\documentclass[journal]{IEEEtran}

\usepackage{xcolor,soul,framed} 

\colorlet{shadecolor}{yellow}
\usepackage[pdftex]{graphicx}
\graphicspath{{../pdf/}{../jpeg/}}
\DeclareGraphicsExtensions{.pdf,.jpeg,.png}

\usepackage[cmex10]{amsmath}
\usepackage{array}
\usepackage{mdwmath}
\usepackage{mdwtab}
\usepackage{eqparbox}
\usepackage{url}
\usepackage{amsthm}
\usepackage{booktabs}
\usepackage{amsfonts}
\usepackage{amssymb}
\usepackage{soul}
\usepackage{amsmath}
\usepackage{bbm}
\usepackage{tikz}
\usepackage{enumitem}
\def\BibTeX{{\rm B\kern-.05em{\sc i\kern-.025em b}\kern-.08em
    T\kern-.1667em\lower.7ex\hbox{E}\kern-.125emX}}
\usepackage{hyperref}
\usepackage{xcolor}
\hypersetup{hidelinks, colorlinks=false, linkcolor=black, urlcolor=black}
\usepackage[short]{optidef}
\usepackage{mathtools}
\usepackage[noadjust]{cite}
\usepackage{textcomp}
\usetikzlibrary{shapes,arrows}
\usepackage{comment}
\usepackage{algorithm}
\usepackage{algpseudocode}
\usepackage{tikz}

\usepackage{MnSymbol}
\usepackage{stmaryrd}
\usepackage{tkz-tab}
\usetikzlibrary{babel}

\newtheorem{theorem}{Theorem} 
\newtheorem*{theorem*}{Theorem}

\newtheorem*{corollary*}{Corollary}

\newtheorem{remark}{Remark}
\newcommand\restr[2]{{
  \left.\kern-\nulldelimiterspace 
  #1 
  \littletaller 
  \right|_{#2} 
  }}
\usepackage{subcaption}
\usepackage{stfloats}  
\usepackage{lipsum}
\newcommand{\littletaller}{\mathchoice{\vphantom{\big|}}{}{}{}}

\usepackage[nolist]{acronym}
    
 \usepackage{bm}

\newcommand{\ma}  [1]{ \bm{#1} } 

\begin{acronym}
\acro{AI}{artificial intelligence}
\acro{LP-WAN}{Low Power Wide Area Networks}
\acro{SEE}{secrecy energy-efficiency} 
\acro{EE}{energy efficiency}
\acro{OMA}{orthogonal multiple access}
\acro{PLS}{Physical Layer Security}

\acro{DSRC}{dedicated short-range communications}
\acro{C-ITS}{cooperative intelligent transport system}
\acro{RSU}{road side unit}
\acro{TDL}{tapped delay line}
\acro{ITS}{Intelligent Transportation Systems}
\acro{IEEE}{Institute of Electrical and Electronics Engineers}
\acro{WAVE}{Wireless Access in Vehicular Environment}
\acro{V2V}{vehicle-to-vehicle}
\acro{V2I}{vehicle-to-infrastructure}
\acro{CCH}{control channel}
\acro{SCH}{service channels}
\acro{STS}{short training symbols}
\acro{LTS}{long training symbols}
\acro{SS}{signal symbol}
\acro{SoA}{state-of-the-art}
\acro{DPA}{data-pilot aided}
\acro{STA}{spectral temporal averaging}
\acro{CDP}{constructed data pilots}
\acro{TRFI}{time domain reliable test frequency domain interpolation}
\acro{MMSE-VP}{minimum mean square error using virtual pilots}
\acro{iCDP}{Improved CDP}
\acro{SBS}{symbol-by-symbol}
\acro{FBF}{frame-by-frame}
\acro{E-TRFI}{Enhanced TRFI}
\acro{SR-CNN}{super resolution CNN}
\acro{DN-CNN}{denoising CNN}
\acro{RBF}{radial basis function}
\acro{CNN}{convolutional neural network}
\acro{TS-ChannelNet}{Temporal spectral ChannelNet}
\acro{WSSUS}{wide-sense stationary uncorrelated scattering}
\acro{TDR}{transmission data rate}
\acro{LSTM}{long short-term memory}
\acro{ALS}{accurate LS}
\acro{SLS}{simple LS}
\acro{ChannelNet}{channel network}
\acro{ADD-TT}{average decision-directed with time truncation}
\acro{WI}{weighted interpolation}
\acro{DD}{decision-directed}
\acro{SR-ConvLSTM}{super resolution convolutional long short-term memory}
\acro{RS}{reliable subcarriers}
\acro{URS}{unreliable subcarriers}
\acro{AE-DNN}{auto-encoder deep neural network}
\acro{AE}{auto-encoder}
\acro{T-DFT}{truncated discrete Fourier transform}
\acro{TA-TDFT}{temporal averaging T-DFT}
\acro{TA}{time averaging}
\acro{PDP}{power delay profile}
\acro{1G}{first generation}
\acro{2G}{second generation}
\acro{3G}{third generation}
\acro{3GPP}{Third Generation Partnership Project}
\acro{4G}{fourth generation}
\acro{5G}{fifth generation}
\acro{6g}{sixth generation}
\acro{B5G}{Beyond 5G}
\acro{802.11}{IEEE 802.11 specifications}
\acro{A/D}{analog-to-digital}
\acro{ADC}{analog-to-digital}
\acro{AM}{amplitude modulation}
\acro{AP}{access point}
\acro{AR}{augmented reality}
\acro{ASIC}{application-specific integrated circuit}
\acro{ASIP}{Application Specific Integrated Processors}
\acro{AWGN}{Additive White Gaussian Noise}
\acro{BCJR}{Bahl, Cocke, Jelinek and Raviv}
\acro{BER}{bit error rate}
\acro{BFDM}{bi-orthogonal frequency division multiplexing}
\acro{BPSK}{binary phase shift keying}
\acro{BS}{base stations}
\acro{CA}{carrier aggregation}
\acro{CAF}{cyclic autocorrelation function}
\acro{Car-2-x}{car-to-car and car-to-infrastructure communication}
\acro{CAZAC}{constant amplitude zero autocorrelation waveform}
\acro{CB-FMT}{cyclic block filtered multitone}
\acro{CCDF}{complementary cumulative density function}
\acro{CDF}{cumulative density function}
\acro{CDMA}{code-division multiple access}
\acro{CFO}{carrier frequency offset}
\acro{CIR}{channel impulse response}
\acro{CM}{complex multiplication}
\acro{COFDM}{coded-\acs{OFDM}}
\acro{CoMP}{coordinated multi point}
\acro{COQAM}{cyclic OQAM}
\acro{CP}{cyclic prefix}
\acro{CR}{cognitive radio}
\acro{CRC}{cyclic redundancy check}
\acro{CRLB}{Cram\'{e}r-Rao lower bound}
\acro{CS}{cyclic suffix}
\acro{CSI}{channel state information}
\acro{CSMA}{carrier-sense multiple access}
\acro{CWCU}{component-wise conditionally unbiased}
\acro{D/A}{digital-to-analog}
\acro{D2D}{device-to-device}
\acro{DAC}{digital-to-analog}
\acro{DC}{direct current}
\acro{DFE}{decision feedback equalizer}
\acro{DFT}{discrete Fourier transform}
\acro{DL}{deep learning}
\acro{DMT}{discrete multitone}
\acro{DNN}{deep neural network}
\acro{FNN}{feed-forward neural network}
\acro{DSA}{dynamic spectrum access}
\acro{DSL}{digital subscriber line}
\acro{DSP}{digital signal processor}
\acro{DTFT}{discrete-time Fourier transform}
\acro{DVB}{digital video broadcasting}
\acro{DVB-T}{terrestrial digital video broadcasting}
\acro{DWMT}{discrete wavelet multi tone}
\acro{DZT}{discrete Zak transform}
\acro{E2E}{end-to-end}
\acro{eNodeB}{evolved node b base station}
\acro{E-SNR}{effective signal-to-noise ratio}
\acro{EVD}{eigenvalue decomposition}
\acro{FBMC}{filter bank multicarrier}
\acro{FD}{frequency-domain}
\acro{FDD}{frequency-division duplexing}
\acro{FDE}{frequency domain equalization}
\acro{FDM}{frequency division multiplex}
\acro{FDMA}{frequency-division multiple access}
\acro{FEC}{forward error correction}
\acro{FER}{frame error rate}
\acro{FFT}{fast Fourier transform}
\acro{FIR}{finite impulse response}
\acro{FM}		{frequency modulation}
\acro{FMT}{filtered multi tone}
\acro{FO}{frequency offset}
\acro{F-OFDM}{filtered-\acs{OFDM}}
\acro{FPGA}{field programmable gate array}
\acro{FSC}{frequency selective channel}
\acro{FS-OQAM-GFDM}{frequency-shift OQAM-GFDM}
\acro{FT}{Fourier transform}
\acro{FTD}{fractional time delay}
\acro{FTN}{faster-than-Nyquist signaling}
\acro{GFDM}{generalized frequency division multiplexing}
\acro{GFDMA}{generalized frequency division multiple access}
\acro{GMC-CDM}{generalized	multicarrier code-division multiplexing}
\acro{GNSS}{global navigation satellite system}
\acro{GS}{guard symbols}
\acro{GSM}{Groupe Sp\'{e}cial Mobile}
\acro{GUI}{graphical user interface}
\acro{H2H}{human-to-human}
\acro{H2M}{human-to-machine}
\acro{HTC}{human type communication}
\acro{I}{in-phase}
\acro{i.i.d.}{independent and identically distributed}
\acro{IB}{in-band}
\acro{IBI}{inter-block interference}
\acro{IC}{interference cancellation}
\acro{ICI}{inter-carrier interference}
\acro{ICT}{information and communication technologies}
\acro{ICV}{information coefficient vector}
\acro{IDFT}{inverse discrete Fourier transform}
\acro{IDMA}{interleave division multiple access}
\acro{IEEE}{institute of electrical and electronics engineers}
\acro{IF}{intermediate frequency}
\acro{IFFT}{inverse fast Fourier transform}
\acro{IoT}{Internet of Things}
\acro{IOTA}{isotropic orthogonal transform algorithm}
\acro{IP}{internet protocole}
\acro{IP-core}{intellectual property core}
\acro{ISDB-T}{terrestrial integrated services digital broadcasting}
\acro{ISDN}{integrated services digital network}
\acro{ISI}{inter-symbol interference}
\acro{ITU}{International Telecommunication Union}
\acro{IUI}{inter-user interference}
\acro{LAN}{local area netwrok}
\acro{LLR}{log-likelihood ratio}
\acro{LMMSE}{linear minimum mean square error}
\acro{LNA}{low noise amplifier}
\acro{LO}{local oscillator}
\acro{LOS}{line-of-sight}
\acro{LP}{low-pass}
\acro{LPF}{low-pass filter}
\acro{LS}{least squares}
\acro{LTE}{Long Term Evolution}
\acro{LTE-A}{LTE-Advanced}
\acro{LTIV}{linear time invariant}
\acro{LTV}{linear time-variant}
\acro{LUT}{lookup table}
\acro{M2M}{machine-to-machine}
\acro{MA}{multiple access}
\acro{MAC}{multiple access control}
\acro{MAP}{maximum a posteriori}
\acro{MC}{multicarrier}
\acro{MCA}{multicarrier access}
\acro{MCM}{multicarrier modulation}
\acro{MCS}{modulation coding scheme}
\acro{MF}{matched filter}
\acro{MF-SIC}{matched filter with successive interference cancellation}
\acro{MIMO}{multiple-input, multiple-output}
\acro{MISO}{multiple-input single-output}
\acro{ML}{machien learning}
\acro{MLD}{maximum likelihood detection}
\acro{MLE}{maximum likelihood estimator}
\acro{MMSE}{minimum mean squared error}
\acro{MRC}{maximum ratio combining}
\acro{MS}{mobile stations}
\acro{MSE}{mean squared error}
\acro{MSK}{Minimum-shift keying}
\acro{MSSS}[MSSS]	{mean-square signal separation}
\acro{MTC}{machine type communication}
\acro{MU}{multi user}
\acro{MVUE}{minimum variance unbiased estimator}
\acro{NEF}{noise enhancement factor}
\acro{NLOS}{non-line-of-sight}
\acro{NMSE}{normalized mean-squared error}
\acro{NOMA}{non-orthogonal multiple access}
\acro{NPR}{near-perfect reconstruction}
\acro{NRZ}{non-return-to-zero}
\acro{OFDM}{orthogonal frequency division multiplexing}
\acro{OFDMA}{orthogonal frequency division multiple access}
\acro{OOB}{out-of-band}
\acro{OQAM}{offset quadrature amplitude modulation}
\acro{OQPSK}{offset quadrature phase shift keying}
\acro{OTFS}{orthogonal time frequency space}
\acro{PA}{power amplifier}
\acro{PAM}{pulse amplitude modulation}
\acro{PAPR}{peak-to-average power ratio}
\acro{PC-CC}{parallel concatenated convolutional code}
\acro{PCP}{pseudo-circular pre/post-amble}
\acro{PD}{probability of detection}
\acro{pdf}{probability density function}
\acro{PDF}{probability distribution function}

\acro{PFA}{probability of false alarm}
\acro{PHY}{physical layer}
\acro{PIC}{parallel interference cancellation}
\acro{PLC}{power line communication}
\acro{PMF}{probability mass function}
\acro{PN}{pseudo noise}
\acro{ppm}{parts per million}
\acro{PRB}{physical resource block}
\acro{PRB}{physical resource block}
\acro{PSD}{power spectral density}
\acro{Q}{quadrature-phase}
\acro{QAM}{quadrature amplitude modulation}
\acro{QoS}{quality of service}
\acro{QPSK}{quadrature phase shift keying}
\acro{R/W}{read-or-write}
\acro{RAM}{random-access memmory}
\acro{RAN}{radio access network}
\acro{RAT}{radio access technologies}
\acro{RC}{raised cosine}
\acro{RF}{radio frequency}
\acro{rms}{root mean square}
\acro{RRC}{root raised cosine}
\acro{RW}{read-and-write}
\acro{SC}{single-carrier}
\acro{SCA}{single-carrier access}
\acro{SC-FDE}{single-carrier with frequency domain equalization}
\acro{SC-FDM}{single-carrier frequency division multiplexing}
\acro{SC-FDMA}{single-carrier frequency division multiple access}
\acro{SD}{sphere decoding}
\acro{SDD}{space-division duplexing}
\acro{SDMA}{space division multiple access}
\acro{SDR}{software-defined radio}
\acro{SDW}{software-defined waveform}
\acro{SEFDM}{spectrally efficient frequency division multiplexing}
\acro{SE-FDM}{spectrally efficient frequency division multiplexing}
\acro{SER}{symbol error rate}
\acro{SIC}{successive interference cancellation}
\acro{SINR}{signal-to-interference-plus-noise ratio}
\acro{SIR}{signal-to-interference ratio}
\acro{SISO}{single-input, single-output}
\acro{SMS}{Short Message Service}
\acro{SNR}{signal-to-noise ratio}
\acro{STC}{space-time coding}
\acro{STFT}{short-time Fourier transform}
\acro{STO}{symbol time offset}
\acro{SU}{single user}
\acro{SVD}{singular value decomposition}
\acro{TD}{time-domain}	
\acro{TDD}{time-division duplexing}
\acro{TDMA}{time-division multiple access}
\acro{TFL}{time-frequency localization}
\acro{TO}{time offset}
\acro{TS-OQAM-GFDM}{time-shifted OQAM-GFDM}
\acro{UE}{user equipment}
\acro{UFMC}{universally filtered multicarrier}
\acro{UL}{uplink}
\acro{US}{uncorrelated scattering}
\acro{USB}{universal serial bus}
\acro{UW}{unique word}
\acro{VLC}{visible light communications}
\acro{VR}{virtual reality}
\acro{WCP}{windowing and \acs{CP}}	
\acro{WHT}{Walsh-Hadamard transform}
\acro{WiMAX}{worldwide interoperability for microwave access}
\acro{WLAN}{wireless local area network}
\acro{W-OFDM}{windowed-\acs{OFDM}}	
\acro{WOLA}{windowing and overlapping}	
\acro{WSS}{wide-sense stationary}
\acro{ZCT}{Zadoff-Chu transform}
\acro{ZF}{zero-forcing}
\acro{ZMCSCG}{zero-mean circularly-symmetric complex Gaussian}
\acro{ZP}{zero-padding}
\acro{ZT}{zero-tail}
\end{acronym}

\usepackage{tikz}
\usetikzlibrary{tikzmark,fit,decorations.pathreplacing}
\usetikzlibrary{calc, positioning, shapes.geometric}
\begin{document}
\bstctlcite{IEEEexample:BSTcontrol}
    \title{
    Explainable Deep Learning for Secrecy Energy-Efficiency Maximization in Ambient
    Backscatter Multi-User NOMA Systems}
    
  \author{Miled~Alam,
      Abdul Karim~Gizzini,~\IEEEmembership{Member,~IEEE, }and
      Laurent~Clavier,~\IEEEmembership{Senior Member,~IEEE}

\thanks{M. Alam and L. Clavier are with IMT Nord Europe, Institut Mines-Télécom, Centre for Digital Systems, F-59653 Villeneuve d'Ascq, France (e-mail: miled.alam@imt-nord-europe.fr, laurent.clavier@imt-nord-europe.fr). A. K. Gizzini is with University of Paris-Est Créteil (UPEC), LISSI/TincNET, F-94400, Vitry-sur-Seine, France.
(e-mail: abdul-karim.gizzini@u-pec.fr). L. Clavier is also with IEMN, UMR CNRS 8520, University of Lille, France.\\
A preliminary version of this work has been presented at the IEEE MeditCom conference~\cite{alam2025secrecy}.

This work was supported by the French National Agency for Research (ANR) via the project n°ANR-22-PEFT-0010 (NF-FOUNDS) of the France 2030 program PEPR réseaux du Futur and COST action CA20120 INTERACT.
}}



\maketitle

\begin{abstract}
In this paper, we investigate the secrecy energy-efficiency (SEE) of a multi-user downlink non-orthogonal multiple access (NOMA) system assisted by multiple ambient backscatter communications (AmBC) in the presence of a passive eavesdropper. We analyze both the trade-off and the ratio  between the achievable secrecy sum-rate and total power consumption. In the special case of two backscatter devices (BDs), we derive closed-form solutions for the optimal reflection coefficients and power allocation by exploiting the structure of the SEE objective and the Pareto boundary of the feasible set. When more than two BDs are present, the problem becomes analytically intractable. To address this, we propose two efficient optimization techniques: (i) an exhaustive grid-based benchmark method, and (ii) a scalable particle swarm optimization algorithm. Furthermore, we design a deep learning-based predictor using a feedforward neural network (FNN), which closely approximates the optimal solutions. Numerical results show that the inclusion of AmBC significantly improves SEE, with gains up to $615\%$ compared to conventional NOMA in high-noise regimes. Additionally, the FNN model achieves more than $95\%$ accuracy compared to the optimal baseline, while reducing complexity. Finally, we employ SHAP (SHapley Additive exPlanations) to interpret the learned model, revealing that the most influential features correspond to the dominant composite channel components, in accordance with the theoretical system model. This demonstrates the potential of explainable artificial intelligence to build trust in energy-efficient and secure AmBC-NOMA systems for next-generation internet of things applications.
\end{abstract}

%
\IEEEpeerreviewmaketitle


\section{Introduction}

\IEEEPARstart{W}{ith} the rapid advancement in wireless communication technologies, the internet of things (IoT) has emerged, interconnecting billions of smart devices worldwide~\cite{yalli2024internet}. As the number of IoT devices continues to grow, significant challenges arise in achieving massive connectivity, extended device lifetimes and high energy efficiency~\cite{you2025next,thomas2025survey}. These growing demands are putting increasing pressure on current network architectures, raising concerns about their ability to meet the rigorous demands of future wireless systems. To efficiently manage the limited radio resources and support the massive number of connected devices, non-orthogonal multiple access (NOMA) has been proposed. In contrast to conventional orthogonal multiple access (OMA), where each user is allocated a distinct time or frequency resource to mitigate interference, NOMA enables multiple users to simultaneously share the same resources by exploiting power-domain multiplexing. At the transmitter, users’ signals are superimposed with different power levels through \textit{superposition coding}. At the receiver, \textit{successive interference cancellation} (SIC) is employed to decode the signals in a hierarchical manner, such that the strongest signal is first decoded by treating the weaker ones as noise and then subtracted from the composite received signal. The procedure continues sequentially until each user’s message is successfully recovered, thereby improving spectral efficiency and resource utilization compared to OMA~\cite{saito2013non,vaezi2019multiple}.  

Although NOMA significantly improves spectral efficiency, energy efficiency remains a major  concern for next-generation IoT networks~\cite{abbasi2020enhancing}. 
Connecting thousands or millions of sensors in a city or a factory, for several years and even decades, require reliability, security and energy efficiency.  This paves the way for the exploration of solutions combining NOMA and ultra-low-power communication technologies, capable of meeting growing sustainability requirements of large-scale IoT deployments.


Ambient backscatter communication (AmBC) presents a promising solution to minimize energy consumption while enhancing communication reliability. It operates by allowing backscatter devices (BDs) to reflect and modulate existing ambient radio frequency (RF) signals, such as those from cellular networks, WiFi, or TV broadcasts, without requiring active RF components or high-power transmitters~\cite{ahmed2024noma}. This enables simultaneous energy harvesting and data transmission, making AmBC a battery-free, low-cost, and spectrally efficient communication paradigm~\cite{chen2024effect}.

Nevertheless, several technical challenges persist, notably the difficulty in separating backscattered signals from ambient ones, as well as increased vulnerability to security threats. Indeed, although AmBC can support reliable wireless communications, the open nature and complexity of the electromagnetic environment make it vulnerable to eavesdropping by malicious actors, posing a serious risk to data confidentiality~\cite{chorti2022context}. As a result, evaluating the secure performance of AmBC-based systems, particularly when integrated with NOMA, is of considerable importance.

In this paper, our main objective is to investigate the secrecy energy-efficiency (SEE) maximization in a multi-user downlink NOMA system assisted by AmBC in the presence of a passive eavesdropper.

Our contributions are manifold and can be summarized as follows:
\begin{itemize}[leftmargin=*, label=\textbf{$\bullet$}]
\item We investigate three practical scenarios for a downlink NOMA-based AmBC system in the presence of a passive eavesdropper: a multi-user system with a single,  two, and  multiple backscatter devices (BDs). The objective in all scenarios is to maximize the SEE defined as both the trade-off and the ratio between the secrecy sum-rate and the power consumption,  subject to constraints on the base station (BS) transmit power, BD reflection coefficients, minimum quality of service (QoS) requirements, and SIC decoding order. The resulting optimization problems are non-convex in all scenarios.
\item  For the  single BD scenario, the solution is based on our previous work~\cite{alam2025secrecy}, where we showed that the problem can be decomposed into two subproblems: a closed-form expression for the optimal reflection coefficient is derived, and the resulting convex power allocation problem is solved analytically.
We then show that our proposed optimal solution can reduce Dinkelbach’s method to a line search when maximizing the ratio secrecy sum-rate vs. power consumption.
\item In the scenario with two BDs, we 
prove that the optimal reflection coefficients lie on the Pareto boundary of the feasible region. Closed-form solutions are obtained for both the optimal reflection vector and the corresponding power allocation while satisfying all practical constraints.
This structure enables efficient maximization of the secrecy sum-rate vs. power consumption ratio through a one-dimensional line search.
\item  In the third scenario
with multiple BDs, the SEE maximization problem becomes analytically intractable due to the non-linear variable coupling and the exponential number of SIC constraints. We address this by proposing two solutions: (i) an exhaustive grid search algorithm for benchmarking purposes, and (ii) a particle swarm optimization (PSO)-based heuristic that directly maximizes the ratio secrecy sum-rate vs. power consumption. 
\item To address the complexity of joint optimization in large-scale scenarios, we design a deep learning-based solution using a feedforward neural network (FNN), which accurately predicts the optimal power allocation and reflection coefficients with low latency. 
\item  Furthermore, we apply SHAP (SHapley Additive exPlanations) to interpret the trained model and quantify the influence of each input feature on the predictions, thereby providing transparency and trust in the AI-driven solution.
\item At last, simulation results validate the effectiveness of the proposed methods. 
Increasing the number of BDs leads to a significant improvement in SEE performance. Moreover, the NOMA-AmBC system consistently outperforms conventional NOMA and OMA schemes, both with and without backscattering, across various deployment scenarios.
\end{itemize}

The remainder of this paper is structured as follows. Section \ref{relatedworks} provides a review of the related literature. Section \ref{sec:sys_mod_pb} introduces the system models and formulates the SEE optimization problems with one or multiple BDs.
Section \ref{section:closedformsolution} presents the proposed solutions. With a single BD, closed-form expressions are derived for the optimal reflection coefficient and the corresponding power allocation. With two BDs,  analytical solutions are also obtained by characterizing the Pareto-optimal reflection coefficients. For the general case with multiple users and multiple BDs, two numerical methods are proposed: an exhaustive grid search for performance benchmarking and a particle swarm optimization (PSO)-based heuristic for low-complexity implementation. Section \ref{DeepLearning} presents an alternative learning-based solution using deep neural networks and interprets the model behavior via SHAP values. Section \ref{sec:res} discusses numerical results to validate the analytical derivations and evaluate the performance under various system parameters. Finally, Section  \ref{Conclusion} concludes the paper and outlines future research directions.
\section{Related Works}\label{relatedworks}
Recently, the integration of ambient backscatter communication (AmBC) and NOMA has attracted increasing attention as a promising strategy to support massive IoT networks. Several studies have examined the combination of AmBC with NOMA, highlighting its advantages in terms of
outage performance~\cite{chrysologou2023performance}, network sum rate~\cite{xie2023bac,raviteja2023fd,rezaei2023beamforming}, and energy efficiency (EE)~\cite{xu2020energy,el2023multi,asif2022energy,ihsan2023energy}. 
A cooperative NOMA framework assisted by backscatter communication is studied in~\cite{xie2023bac}, where the authors focus on enhancing spectral efficiency. They formulate an optimization problem and derive a closed-form expression for the optimal backscattering coefficient, leading to a significant gain in the system's achievable sum rate.
To further improve performance,~\cite{raviteja2023fd} proposes an iterative algorithm that jointly optimizes power allocation and reflection coefficients in an AmBC-enabled downlink system, aiming to maximize the sum rate under a perfect SIC assumption.

Beyond reliability and spectral efficiency, EE has also emerged as a critical performance metric in the design of NOMA-assisted AmBC systems, particularly for energy-constrained IoT applications. In~\cite{xu2020energy}, the authors addressed the non-convex EE maximization problem in an AmBC network for a two-user downlink NOMA system using an iterative method. 
As a further advancement, the achievable rate of AmBC networks was derived in~\cite{el2023multi} and a joint optimization framework was proposed to maximize the EE of multi-user AmBC networks, while also studying the impact of imperfect channel state information (CSI) on their solution. 
In a more complex scenario involving cognitive radio, the authors of~\cite{huang2025research} tackled the joint optimization of power allocation and reflection coefficients in a downlink NOMA-AmBC cognitive network with multiple BDs. They formulated a non-convex problem aimed at maximizing EE and proposed a two-stage approach based on Lagrangian optimization and particle swarm optimization. Simulation results demonstrate that their proposed scheme significantly outperforms conventional approaches in terms of EE.

However, the backscattered signal may also be vulnerable to eavesdropping due to the inherently open and broadcast nature of wireless communication channels~\cite{saad2014physical}. This raises significant concerns regarding the security of NOMA-assisted AmBC systems, especially in scenarios involving sensitive data or critical IoT applications. As a result, physical layer security (PLS) has emerged as a promising solution to safeguard information by exploiting the physical characteristics of the wireless medium, without relying on higher-layer cryptographic methods. Recent research has begun to explore the integration of PLS techniques into NOMA-based AmBC architectures to enhance the system’s resilience against eavesdropping attacks. 
To mitigate this vulnerability, the authors in~\cite{pei2024secrecy} proposed the use of reconfigurable intelligent surfaces (RIS) to enhance the PLS of AmBC networks. Specifically, they focused on maximizing the secrecy outage probability and derived a closed-form solution to characterize system performance under the considered scenario.
In~\cite{jia2023secrecy}, the secrecy performance of AmBC-enabled smart transportation networks was studied, with jamming used to disrupt eavesdropper reception. 
The work of~\cite{li2023physical} examined the impact of decode-and-forward relays on the secrecy performance of AmBC networks.
In~\cite{khan2023joint}, the authors consider a two-user downlink NOMA network with AmBC in the presence of multiple eavesdroppers. Their focus is on maximizing the secrecy rate of the near user, subject to the reflection power of the backscatter device and the transmit power of the BS. The problem is effectively solved using a low-complexity iterative method based on the Lagrangian dual approach.
In~\cite{huang2025physical}, a cooperative NOMA-AmBC system with an optimal tag selection strategy was proposed to enhance both reliability and security. Closed-form expressions for outage and intercept probabilities were derived, considering non-collusive and collusive eavesdroppers. The results show that tag selection improves reliability, while the presence of collusive eavesdroppers significantly increase the risk of interception. In~\cite{pei2025physical}, the PLS of AmBC-NOMA networks was studied in the presence of randomly distributed non-collusive eavesdroppers. The authors proposed using artificial noise and an eavesdropper-exclusion zone to enhance security, and derived outage and intercept probabilities to characterize the reliability-security trade-off.

To sum up, the aforementioned studies on NOMA-AmBC systems have investigated key performance metrics such as outage probability, sum rate, and secrecy rate under various system assumptions and configurations. To the best of our knowledge, a rigorous investigation of the trade-off between secrecy and energy consumption is still lacking, leaving the secrecy energy-efficiency performance largely unexplored in this context.

The closest related works are our two previous studies~\cite{savard2024secrecy} and~\cite{alam2025secrecy}. To the best of our knowledge, the study presented in~\cite{alam2025secrecy} was the first to address the secrecy energy-efficiency optimization problem in AmBC systems. In that work, we considered a single backscatter device (BD) and focused on the joint optimization of the reflection coefficient and power allocation, for which closed-form solutions were derived.
Moreover, the work in~\cite{alam2025secrecy} can be seen as an extension of our earlier study~\cite{savard2024secrecy}, where a conventional NOMA system without backscatter was considered. By incorporating a BD in~\cite{alam2025secrecy}, we demonstrated a notable improvement in SEE compared to the system in~\cite{savard2024secrecy}.

In this paper, we extend our previous study to scenarios with multiple BDs. We first examine the case with two BDs, where closed-form solutions are obtained for the Pareto-optimal reflection coefficients and the corresponding power allocation. We then consider the general case with multiple BDs, where the SEE maximization problem becomes analytically intractable; for this setting, we propose an exhaustive grid-search benchmark and a particle swarm optimization (PSO)-based heuristic. The SEE objective is defined as both the trade-off and the ratio between the secrecy sum-rate and the power consumption.

\section {System Model and Problem Formulation}\label{sec:sys_mod_pb}

This section present the considered system models as well as the secrecy energy-efficiency optimization problem with one or multiple BDs.

\subsection{Multi-User NOMA with a Single Backscatter Device}
\subsubsection{\bf{System Model}}\label{sectionsystemmodel1}
\begin{figure}[t]
\centering
\includegraphics[width=0.8\columnwidth, height=0.2\textheight]{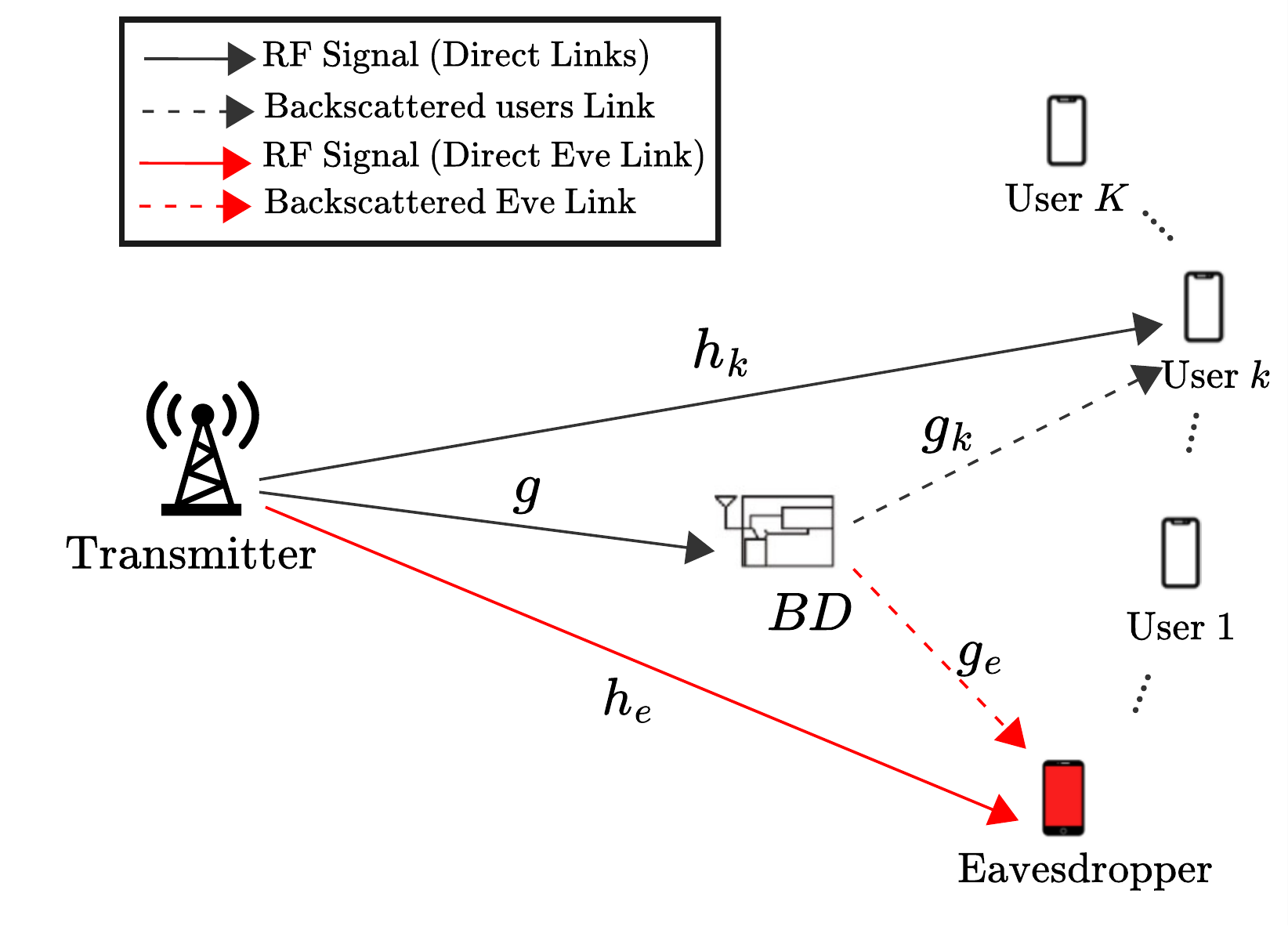}
\caption{Downlink multi-user NOMA system assisted by ambient backscatter with an eavesdropper.\label{fig:sys_mod}}
\end{figure}

As shown in Fig.~\ref{fig:sys_mod}, we consider a downlink multiple access NOMA communication system consisting of a single transmitter or source (e.g., \ac{BS}, WiFi hotspot, etc.), an arbitrary number of legitimate users $K \ge 2$ (e.g., IoT devices, smartphones, etc.), and a single eavesdropper. All nodes are equipped with a single antenna. The system is assisted by an ambient backscatter device (BD) capable of harvesting energy and use it to reflect the radio frequency (RF) signal while superposing its own low-rate information.
The transmitter employs \textit{superposition coding} and broadcasts the signal $X = \sum_{k=1}^K \sqrt{p_{k}} X_{k}$, where $X_k$ denotes the message intended for user $k \in \llbracket 1,K \rrbracket$, satisfying $\mathbb{E}[|X_k|^2]=1$, and $p_k$ represents the power allocated to user $k$, subject to the transmitter power constraint $\sum_{k=1}^K p_k \leq P_{\max}$. 
During the NOMA transmission, both the BD and the eavesdropper receive the superimposed signal from the BS. 
We consider a worst-case scenario in which the eavesdropper is capable of intercepting both the direct transmission from the BS and the backscattered signal. 
Given the low-rate IoT context considered in this study, we assume a narrowband transmission with sufficiently long symbol durations, allowing us to neglect both processing delays and propagation effects~\cite{khan2023joint, kang2018riding}. Furthermore, we assume that the BD transmits at a much lower rate than the IoT device, such that each backscatter symbol (taking the value $+1$ or $-1$) remains constant over a sufficiently long sequence of IoT symbols. This assumption ensures that the phase of the reflected signal does not change within the duration of an IoT symbol, thereby avoiding additional phase variations that would otherwise need to be modeled.

The system model under consideration is applicable to various real-world scenarios in emerging wireless networks, especially those requiring low-power communications with enhanced security. For  instance, in smart environments such as intelligent buildings or industrial facilities, a central base station (e.g., Wi-Fi access point or 5G transmitter) can simultaneously serve multiple IoT devices using NOMA. At the same time, passive sensors equipped with backscatter capabilities can transmit their own low-rate information (e.g., temperature, humidity, or motion detection) by reflecting ambient RF signals from the main transmitter, without requiring any active RF components~\cite{mondal2025comprehensive}. This makes AmBC particularly attractive for batteryless or ultra-low-power devices in the IoT.

The received signals at the $k$-th legitimate user and at the eavesdropper are composed of both the direct signal transmitted from the source and the backscattered signal, expressed as follows
\begin{equation}
\begin{aligned}
Y_{k} & \quad= \quad  \tikzmark{startk}h_{k} X\tikzmark{endk} \quad + \quad \tikzmark{startk1} {\sqrt{\rho} g g_{k} B X}\tikzmark{endk1} \quad+\quad  Z_{k},  \\
Y_{e} &\quad =\quad  \tikzmark{starte}h_{e} X\tikzmark{ende} \quad + \quad \tikzmark{starte1} {\sqrt{\rho} g g_{e} B X} \tikzmark{ende1} \quad +\quad  Z_{e},
\end{aligned}
\end{equation}

\begin{tikzpicture}[remember picture, overlay]
\node[align=center] at ([yshift=-0.8cm]$(pic cs:startk)!0.5!(pic cs:ende)$) {\scriptsize \textit{Transmitter Signal}};
\node[align=center] at ([yshift=-0.8cm]$(pic cs:startk1)!0.5!(pic cs:ende1)$) {\scriptsize \textit{Backscattered Signal}};
\draw[red] ([shift={(-0.1cm,0.4cm)}]pic cs:startk) rectangle ([shift={(0.1cm,-0.3cm)}]pic cs:ende);
\draw[red] ([shift={(-0.1cm,0.4cm)}]pic cs:startk1) rectangle ([shift={(0.1cm,-0.3cm)}]pic cs:ende1);
\end{tikzpicture}\\
\\
where $h_k$, $h_e$, $g$, $g_k$ and $g_e$ represent the channel coefficients\footnote{The channel coefficients are based on a simplified path loss model, as justified by the short-range nature of AmBC~\cite{el2021energy,chen2021backscatter,nazar2021ber}; further elaboration can be found in Section~\ref{sec:res}.}  from source to user $k$, from source to eavesdropper, from source to BD, from BD to user $k$ and from BD to eavesdropper, respectively --- see Fig. \ref{fig:sys_mod}. The term $\rho\in [0,1]$ is the reflection coefficient of the BD. The quantities $Z_{k}\sim \mathcal{N}\left(0,\sigma_{k}^{2}\right)$ and $Z_{e}\sim \mathcal{N}\left(0,\sigma_{e}^{2}\right)$ are the additive white Gaussian noises (AWGN) at the $k$-th user and the eavesdropper respectively, and $B$ is the signal transmitted by the BD. 
We assume $B = 1$, which simplifies the model by treating the BD as a passive relay. This assumption does not imply any loss of generality, provided that the data rate of the BD is significantly lower than that of the BS. Under this condition, $B$ can be considered constant over the entire backscattered packet. For binary signaling, setting $B = 1$ is both valid and sufficient~\cite{el2021energy}. It is worth noting that the harvested energy at the BD is not explicitly considered in our analysis~\cite{el2023multi,zhou2019ergodic,khan2021energy}.

In addition to the backscattering model assumptions, channel estimation represents another critical aspect of our system. Channel estimation can be effectively accomplished through training methods; however, obtaining perfect channel state information (CSI) remains challenging in practical implementations due to estimation errors. To focus on the secrecy-energy efficiency and facilitate the mathematical analysis, we make the assumption that perfect CSI is available at all nodes, which aligns with approaches taken in related studies~\cite{nazar2021ber,li2022effective}. Analysis of scenarios involving imperfect CSI is left for future research.
Throughout this study, we assume normalized channel gains defined as
$H_{k}^{(0)}$ = ${h_{k}^{2}}/{\sigma_{k}^{2}}$, $H_{k}^{(1)}(\rho) = \left(h_{k} + \sqrt\rho gg_{k}\right)^{2}/{\sigma_{k}^{2}}$, $k \in\{ \llbracket 1,K\rrbracket,e$\}, where $H_{k}^{(0)}$ and $H_{k}^{(1)}(\rho)$ represent the normalized channel gains with respect to the noise variance when the backscatter device is either not reflecting any signal ($B=0$) or reflecting the signal from the transmitter ($B = 1$), respectively. The case $B=0$ corresponds to the baseline system model studied in \cite{savard2024secrecy}, where no backscattered signal interferes with the direct link.\\
Without loss of generality, we assume that the channel gains are ordered as follows 
\begin{equation}
    H_{1}^{(0)} \leq \cdots \leq H_{M-1}^{(0)} \leq H_{e}^{(0)} < H_{M}^{(0)} \leq \cdots \leq H_{K-1}^{(0)} < H_{K}^{(0)},
    \label{Channel order}
\end{equation}
i.e. $M-1$ of the $K$ legitimate users experience channel gains lower than those of the eavesdropper.

In power domain NOMA, the transmitter employs a superposition coding scheme, where each legitimate user $k$ decodes their intended message via successive interference cancellation (SIC)~\cite{vaezi2019noma,makki2020survey}. 
It is worth emphasizing that the SIC decoding order is established based solely on the direct channels between the transmitter and the users, without considering the backscattered path. This is because the behavior of the BD is not controlled by the transmitter; it operates opportunistically and independently when modulating its own information. In this decoding process, user $k$ first decodes the signals intended for users with weaker direct channel gains, indexed by  $1 \leq \ell \leq k-1$, while treating the signals from users with stronger channels, indexed by $k+1 \leq \ell \leq K$, as interference during the decoding of its own message.
Accordingly, the signal-to-interference-plus-noise ratio (SINR) required for user $k$ to successfully decode their message at receiver $i$ is given by 
\begin{equation}\label{SINR}
    \gamma_{k \to i} = \frac{H_{i}^{(1)}(\rho) p_k}{H_{i}^{(1)}(\rho)(p_{k+1}+\ldots+p_{K}) + 1} .
\end{equation}

To ensure that the SIC decoding process proceeds correctly according to the ascending  order of channel gains in~\eqref{Channel order}, the following set of SINR constraints must be satisfied:
\begin{equation}\label{constraint}
    \gamma_{k \to i} \geq \gamma_{k \to k}, \quad \forall\, k < K,\ \forall\, i \geq k+1.
\end{equation}
These constraints guarantee that any receiver $i$ (with stronger channel conditions than user $k$) can successfully decode and cancel the interference from user~$k$'s signal before decoding its own message.

Under the satisfaction of these conditions, the achievable rate $R_k$ for decoding the message intended for user~$k$ at its corresponding receiver is expressed as follows~\cite{el2021energy,cover1999elements}
\begin{equation}\label{achievablerateusers}
    R_k(\rho, \mathbf{p}) = \frac{1}{2} \log_2 \left(1 + \min_i \gamma_{k \to i} \right) = \frac{1}{2} \log_2 \left(1 + \gamma_{k \to k} \right).
\end{equation}
The equality holds due to the fact that user~$k$ experiences the weakest SINR among all the users decoding its message, hence the minimum SINR is attained at its own receiver.

In the considered system model, the eavesdropper is assumed to be part of the user group served by the source. However, in contrast to the legitimate users, the eavesdropper conducts a passive attack, remaining undetected by the transmitter. This stealthy behavior allows the eavesdropper to gain knowledge of the global channel gain ordering, and thus to infer the SIC decoding sequence applied by the legitimate users.

As a result, the eavesdropper is capable of attempting to decode the message intended for any user $k$ by exploiting the superimposed nature of the transmitted signal. The achievable rate at the eavesdropper for decoding user $k$’s message is thus expressed as
\begin{equation}\label{achievableeav}
    R_{k}^{e}\left(\rho, \mathbf{p}\right) = \frac{1}{2} \log_{2} \left(1 + \gamma_{k \to e} \right),
\end{equation}
where $\gamma_{k \to e}$ denotes the SINR at the eavesdropper when attempting to decode the signal intended for user~$k$.

 Throughout this study, the network is designed to operate under a global power limitation, denoted by $P_{\max}$, ensuring that the total transmit power allocated across all users satisfies the condition $\sum_{k} p_k \leq P_{\max}$. Additionally, each user is subject to a minimum quality of service (QoS) constraint, requiring that its achievable data rate fulfills $R_k(\rho,\mathbf{p}) \geq R_{\min, k}$, $k \in \llbracket 1, K \rrbracket$.

Throughout the remainder of the paper, we introduce the following notation to simplify the derivations:
\begin{equation}\label{notation}
A_{k} \!=\! 2^{2R_{\min,k}}, \theta_{k}(\mathbf{{p}}) \!=\! \sum\limits_{i=k}^{K} p_{i},\!\!\ k \in \llbracket 1,K\rrbracket  \text{ and }  \theta_{K+1}(\mathbf{{p}})\!=\!0.
\end{equation}
All the aforementioned conditions lead to the feasible set, denoted as
\begin{align}\label{feasabileset}
    &\mathbf{\Pi} = \Bigg\{ (\rho, \mathbf{p}) \in [0, 1] \times \mathbb{R}_+^K \mid 0 \leq \rho \leq 1, \ \theta_{1}(\mathbf{p}) \leq P_{\max},\nonumber \\
    & R_k(\rho, \mathbf{p}) \geq R_{\min, k}, \gamma_{k \to i} \!\geq\! \gamma_{k \to k}, \forall 1 \leq k \leq K, \forall i \ge k+1 \!\Bigg\}.
\end{align}
\subsubsection{\textbf{Problem Formulation}}
In the field of physical layer security, the goal is to leverage the imperfections in the communication channels, such as noise, to differentiate the signal quality between legitimate users and the eavesdropper. For secure communication to be feasible, the eavesdropper must experience a weaker channel compared to the legitimate receivers. The secrecy rate quantifies this concept, representing the difference in data rates between the legitimate user and the eavesdropper when attempting to decode the same message. For each user $k\in\llbracket 1,K \rrbracket$, the achievable secrecy rate is defined as~\cite{savard2024secrecy},~\cite{yao2019secrecy}    
\begin{equation}\label{achievablesecrecyrate}  R_{k}^{s}\left(\rho,\mathbf{p}\right) =  \left[R_{k}\left(\rho,\mathbf{{p}}\right) - R_{k}^{e}\left(\rho,\mathbf{{p}}\right)\right]^{+}, 
\end{equation}
where $[x]^+=\max\{0;x\}$.\\
Considering the channel ordering defined earlier in \eqref{Channel order}, as well as the SINR constraint described in \eqref{constraint}, it follows that the secrecy rate for the \( M-1 \) weakest users will be zero. In contrast, users indexed from \( M \) to \( K \) achieve strictly positive secrecy rates, due to their sufficiently strong channel conditions that enable secure communication.

Based on the previous analysis, the secrecy energy-efficiency (SEE)  optimization problem can be expressed as a scalarized objective that jointly maximizes the total achievable secrecy sum-rate while minimizing the overall power consumption. The optimization problem is formulated as follows
\begin{align*}
\textbf{(SEE1)}&~ \max_{\rho, \mathbf{{p}} } 
\left( \sum_{k=1}^{K} R_{k}^{s}\left(\rho, \mathbf{{p}}\right) - \alpha \big( \theta_{1}(\mathbf{p}) + P_c \big) \right), \\
\text{ s.t. } 
& \theta_1(\mathbf{{p}}) \leq P_{\max} \tag{C1} \label{eq:C1}, \\
& \theta_k(\mathbf{{p}}) \geq A_k \theta_{k+1}(\mathbf{{p}}) + \frac{A_k - 1}{H_{k}^{(1)}(\rho)}, \quad k \in \llbracket 1, K \rrbracket \tag{C2} \label{eq:C2}, \\
& \gamma_{k \to i} \geq \gamma_{k \to k}, \quad \forall k < K, \ \forall i \geq k+1 \tag{C3} \label{eq:C3}, \\
& 0 \leq \rho \leq 1, \tag{C4} \label{eq:C4}
\end{align*}
where $P_c$ represents the constant circuit power consumption.
 The parameter $\alpha > 0$ governs the balance between maximizing the total achievable secrecy rate and minimizing power consumption as in~\cite{savard2024secrecy,el2023multi}. Larger values of $\alpha$ lead to an optimization focused on maximizing the secrecy sum-rate, while smaller values shift the focus toward reducing the overall power consumption.
The constraints are defined as follows: (C1) limits the total transmission power within the maximum budget allowed at the transmitter, (C2) ensures that each user's QoS requirement is satisfied, which is obtained by rewriting the $R_k(\rho,\mathbf{p}) \geq R_{\min,k}$, (C3) imposes the decoding order required by the SIC mechanism by maintaining sufficient SINR separation, and (C4) enforces the reflection coefficient to lie within its feasible range between 0 and 1.

\subsection{Multi-User NOMA with Multiple Backscatter Devices}
\begin{figure}[t]
\centering
\includegraphics[width=0.9\columnwidth, height=0.2\textheight]{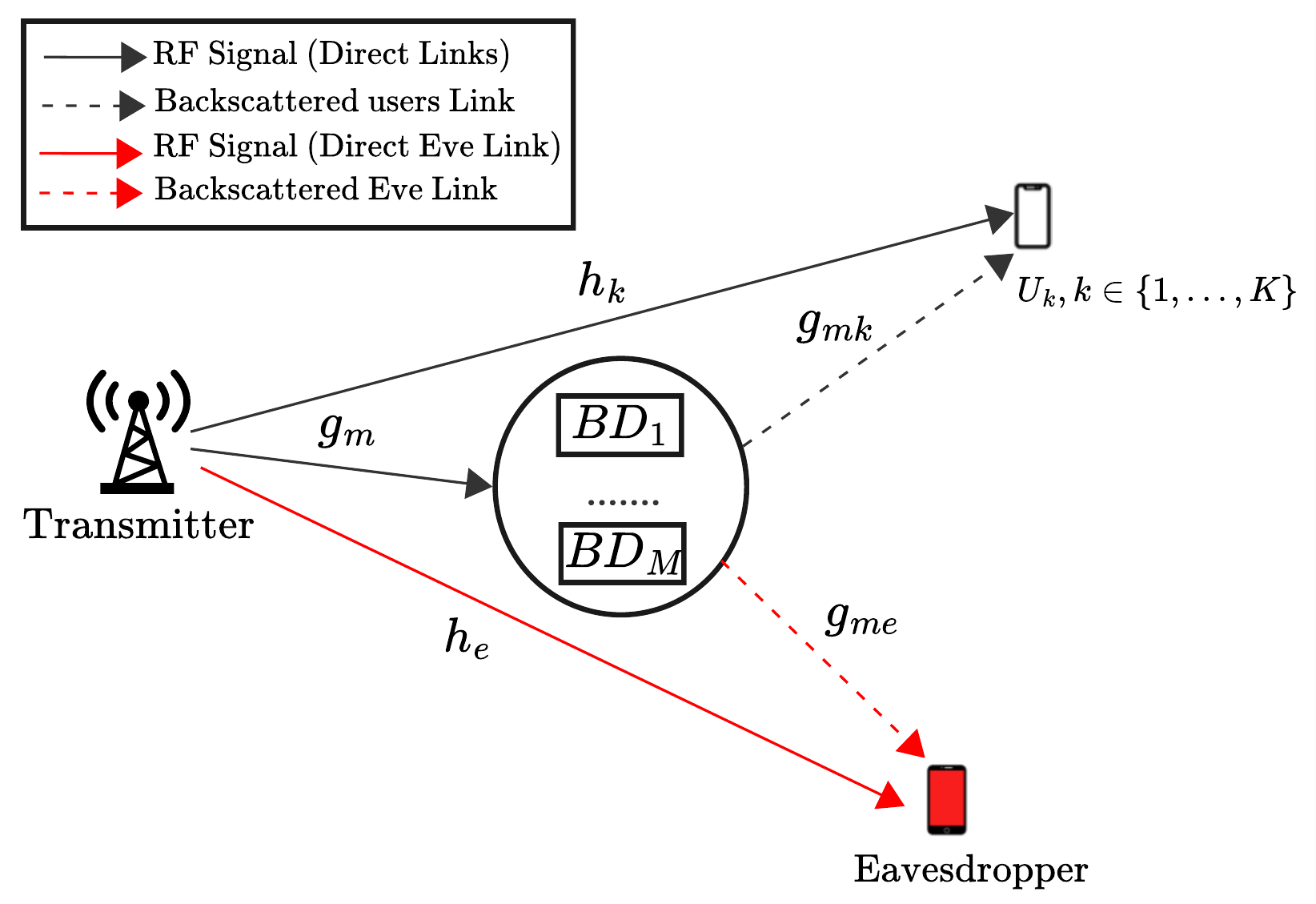}
\caption{Multi-user NOMA system with multiple ambient backscatter devices and a single eavesdropper.\label{fig:sys_mod2BD}}
\end{figure}

In this section, we extend our study to a generalized multi-user NOMA system incorporating multiple ambient backscatter devices (BDs), as depicted in Fig.~\ref{fig:sys_mod2BD}. The system includes $ K $ legitimate users, $ M $ BDs, and a single passive eavesdropper. While this configuration provides a comprehensive model for evaluating both performance and security, the joint optimization of the \( M \) reflection coefficients is highly non-linear and becomes analytically intractable as the system dimensions grow, as discussed in Section~\ref{subsubsec:GeneralCase}. To gain analytical insight before addressing the full complexity of the general case, we first consider a simplified but representative
configuration with $ K = 2 $ users and $ M = 2 $ BDs. This simplified yet non-trivial scenario allows us to derive closed-form expressions, enabling a tractable performance analysis while capturing the key challenges associated with multiple BDs.
\subsubsection{\textbf{System Model} (Case  $K=2 $, $ M=2 $)}
The overall transmission structure remains aligned with the setup of Section~\ref{sectionsystemmodel1}. The transmitter (e.g., BS, etc.) transmits a superimposed signal $X = \sqrt{p_{1}}X_{1} + \sqrt{p_{2}}X_{2} $ where $X_k$ denotes the unit-power message intended for user $U_{k}$, and $p_{k}$ is the power allocated to that user, satisfying $p_{1} + p_{2} \leq P_{\max}$  to both users, $U_{1}$ and $U_{2}$, while the communication is overheard by a passive eavesdropper. Unlike the single BD scenario, the current configuration includes two BDs, denoted $BD_{1}$ and $BD_{2}$,  each reflecting the incident signal with an reflection coefficient $\rho_{1}$ and $\rho_2$, respectively.  This configuration, depicted in Fig.~\ref{fig:sys_mod2BD}, provides a relevant and analytically manageable scenario that allows us to investigate the benefits and limitations of incorporating multiple BDs from both performance and physical layer security perspectives.

Each receiver thus observes a composite signal comprising the direct transmission and the contributions from both backscattering paths. Using the previously introduced notations, the received signal at user $U_k$, $k \in \{1,2\}$, writes as
\begin{equation}
Y_{k} = \Big(h_{k}  + \sqrt{\rho_{1}}g_{1}g_{1k} + \sqrt{\rho_{2}} g_{2}g_{2k}\Big) X + Z_{k},
\end{equation}
and the eavesdropper receives
\begin{equation}
Y_{e} =\Big( h_{e}  + \sqrt{\rho_{1}}g_{1}g_{1e} + \sqrt{\rho_{2}} g_{2}g_{2e}\Big) X + Z_{e},
\end{equation}
where \( h_k \) and \( h_e \) are the direct channels from the transmitter to user \( k \) and to the eavesdropper, respectively. The terms \( g_1 \) and \( g_2 \) represent the channels from the transmitter to backscatter devices \( BD_1 \) and \( BD_2 \), whereas \( g_{mk} \) and \( g_{me} \) correspond to the links from backscatter device \( BD_m \) to user \( k \) and to the eavesdropper, respectively. Finally, \( Z_k \) and \( Z_e \) denote independent additive white Gaussian noise (AWGN) terms at user \( k \) and at the eavesdropper, respectively.

Accordingly, the effective normalized channel gain at any node \( k \in \{1,2,e\} \) in the presence of two BDs is given by
\begin{equation}\label{normalizedchannelgain2BD}
    H_{k}^{(2)}(\bm\rho) = \frac{\left(h_{k} + \sqrt{\rho_{1}}\,g_{1}g_{1k} + \sqrt{\rho_{2}}\,g_{2}g_{2k}\right)^2}{\sigma_{k}^{2}},
\end{equation}
where \( \bm \rho = (\rho_{1}, \rho_{2}) \) denotes the vector of reflection coefficients associated with \( BD_1 \) and \( BD_2 \), respectively. To simplify the subsequent expressions, we introduce the following normalized channel notation:
\begin{equation}\label{notation2}
    G_{k} = \frac{h_{k}}{\sigma_{k}},\    
    G_{1k} = \frac{g_{1}g_{1k}}{\sigma_{k}},\   
    G_{2k} = \frac{g_{2}g_{2k}}{\sigma_{k}}, \ \forall k \in \{1,2,e\}.
\end{equation} 

Using the above notation, the signal-to-interference-plus-noise ratio (SINR) for user $k$ to successfully decode their message at receiver $i$ becomes
\begin{equation}\label{SINR2BD}
    \gamma_{k \to i} = \frac{H_{i}^{(2)}(\bm{\rho}) p_k}{H_{i}^{(2)}(\bm{\rho})(p_{k+1}+\ldots+p_{K}) + 1} .
\end{equation}
Consequently, the achievable rate expressions for the  legitimate users and the eavesdropper follow directly from~\eqref{achievablerateusers} and~\eqref{achievableeav}, while the achievable secrecy rate is still defined as in~\eqref{achievablesecrecyrate}, subject to the substitution:
$H_{k}^{(1)}(\rho) \rightarrow H_{k}^{(2)}(\bm \rho)$.

\subsubsection{\textbf{Problem Formulation}}
In line with the SEE metric introduced previously, we formulate the optimization problem for this setting as follows
\begin{align*}
\textbf{(SEE2)}&~ \max_{\bm \rho, \mathbf{{p}}} 
\left( \sum_{k=1}^{2} R_{k}^{s}\left(\bm \rho, \mathbf{{p}}\right) - \alpha \left( \sum_{k=1}^{2} p_k + P_c \right) \right), \\
\text{ s.t. } 
& \theta_1(\mathbf{{p}}) \leq P_{\max} \tag{C1} \label{eq:C12BD}, \\
& \theta_k(\mathbf{{p}}) \geq A_k \theta_{k+1}(\mathbf{{p}}) + \frac{A_k - 1}{H_{k}^{(2)}(\bm \rho)}, \quad \forall k \in \{ 1, 2\} \tag{C2} \label{eq:C2BD}, \\
& \gamma_{1 \to 2} \geq \gamma_{1 \to 1}, \tag{C3} \label{eq:C32BD} \\
& 0 \leq \rho_{1} , \rho_{2} \leq 1. \tag{C4} \label{eq:C42BD}
\end{align*}

The resolution of this problem is presented in section~\ref{section:closedformsolution} , which closed form solution is presented.

\section{Secrecy Energy Efficiency Maximization via Conventional Approach}\label{section:closedformsolution}
In this section, we present the  optimal solutions to problems \textbf{(SEE1)} and \textbf{(SEE2)}. Both  problems are inherently non-convex due to variable coupling, which makes direct solutions intractable. This issue is similar to that studied in \cite{khan2023joint}, where the secrecy rate is maximized for two users in the presence of multiple eavesdroppers and a BD, using a duality-based iterative algorithm to address the coupling between optimization variables. In contrast, our approach reveals that the trade-off between maximizing the secrecy rate and minimizing energy consumption can be solved in closed-form, thus eliminating the need for iterative optimization in both scenarios. 
\subsection{Closed-Form Secrecy Energy-Efficiency Solution for Multi-User NOMA with a Single Backscatter Device}
Building on the methodology presented in \cite{el2021energy,xu2020energy}, we demonstrate that the optimization problem \textbf{(SEE1)} can be solved efficiently by decoupling it into two subproblems, while preserving optimality. Specifically, (\romannumeral 1) we first optimize the reflection coefficient (RC) \(\rho\) for an arbitrary power allocation \(\mathbf{p}\); (\romannumeral 2) we then optimize the power allocation \(\mathbf{p}\) while keeping the RC \(\rho\) fixed at its optimal value.
\subsubsection{\textbf{Optimal Reflection Coefficient}}
To start, we fix an arbitrary power allocation $ \mathbf{p} \in \Pi $ and focus on solving the optimization problem \textbf{(SEE1)} with respect to the RC \( \rho \). Interestingly, the optimal RC does not depend on the power allocation \( \mathbf{p} \), and it can be expressed in a closed form as 
 \begin{equation}\label{eq:optimal RC}
        \rho^{*} = 
        \begin{cases} 
            1, & \text{ if } \  \frac{gg_{k+1}}{\sigma_{k+1}} > \frac{gg_{k}}{\sigma_{k}}, \forall k < K,\\
            0, & \text{otherwise}.
        \end{cases}
    \end{equation}

The proof is similar to that presented in our previous work~\cite{alam2025secrecy}.
\subsubsection{\textbf{Optimal Power Allocation}}
Once the optimal RC \( \rho^{*} \) is selected as in~\eqref{eq:optimal RC}, constraints (C3) and (C4) are inherently satisfied. As demonstrated in~\cite{alam2025secrecy}, this choice leads to the following ordering of the effective channel gains:
$H_{1}^{(1)}(\rho^{*}) \leq \cdots \leq H_{M-1}^{(1)}(\rho^{*}) \leq H_{e}^{(1)}(\rho^{*}) < H_{M}^{(1)}(\rho^{*}) \leq \cdots \leq H_{K-1}^{(1)}(\rho^{*}) < H_{K}^{(1)}(\rho^{*})$, 
which guarantees the feasibility of the SIC decoding structure. Under this channel ordering and for a fixed value of \( \rho^{*} \), the original problem \textbf{(SEE1)} simplifies to a pure power allocation problem, i.e.,
\label{SEE2}
\begin{align*}
\textbf{(SEE3)}&~ \Psi_{SEE}(\mathbf{p})\! =\!  \max_{\mathbf{{p}}} \left( \sum_{k=1}^{K} R_{k}^{s}\left(\rho^*, \mathbf{{p}}\right) \!-\! \alpha \big( \theta_{1}(\mathbf{p}) + P_c \big) \right), \\
\text{ s.t. } 
&\theta_{1}(\mathbf{{p}})  \leq P_{\max} \tag{C1'} \label{eq:C1'},\\
   &\theta_{k}(\mathbf{{p}}) \geq A_k \theta_{k+1}(\mathbf{{p}}) + \frac{A_{k} - 1}{H_{k}^{(1)}(\rho^{*})},\   k \in \llbracket 1,K\rrbracket, \tag{C2'} \label{eq:C2'} 
\end{align*}

The residual optimization problem admits a convex formulation, since: (i) the objective function is strictly concave with respect to \( \mathbf{p} \), as established through a negative definite Hessian analysis in Appendix B of~\cite{savard2024secrecy}; and (ii) the constraints (C1')–(C2') are linear, thereby defining a convex feasible set.

\label{feasiblecondition}
\textit{\textbf{Feasibility condition}}: Given the constraints mentioned earlier, the feasible set of the optimization problem \textbf{(SEE3)} can be expressed as
\begin{equation}
\hspace{-0.08cm}\left\{\! \mathbf{p}\!\in \!\mathbb{R}_{+}^{K} \bigg|  \theta_{1}(\mathbf{p}) \! \leq \! P_{\max},\, \theta_{k}(\mathbf{p}) \!\ge\! A_{k} \theta_{k+1}(\mathbf{p})\! +\! \frac{A_{k}\!-\!1 }{H_{k}^{(1)}(\rho^{*})} \!\!\right\}.
\label{eq:feasible set}
\end{equation}
It is important to note that, depending on the channel conditions, the available power budget \(P_{\max}\) may not be sufficient to satisfy all users’ QoS requirements in (C2’). The problem is feasible only if \(P_{\max}\) is greater than or equal to a minimum required power \(P_{\min}\), which represents the total power needed to satisfy all QoS constraints simultaneously. Formally, it can be expressed as: $P_{\min} \coloneqq \sum_{k=1}^{K} \frac{A_{k} - 1}{H_{k}^{(1)}(\rho^{*})} \prod_{j=1}^{k-1} A_{j}.$ Therefore, a feasible solution exists if and only if $P_{\max} \geq P_{\min},$ which is consistent with the results in our previous work on optimal power allocation policies for \(K\)-receiver downlink NOMA with an eavesdropper, excluding the backscatter device~\cite{savard2024secrecy}. 

If the aforementioned feasible set is non-empty, and since \textbf{(SEE3)} is a convex optimization problem, it therefore admits a single global optimal solution. This optimal power allocation is obtained through the application of the Karush-Kuhn-Tucker (KKT) optimality conditions.
As proven in Appendix B of \cite{savard2024secrecy}, the expressions of the optimal power $p_{k}^{*}(\alpha)$, for  $k \in \llbracket 1, K-1 \rrbracket$, allocated to all users except the one with the strongest channel gain,  are functions of the optimal power allocated to the strongest user  $p_{K}^{*}(\alpha)$, turning the multi-variable optimization problem under consideration into a single variable one, which is proven to be a concave one,  yielding the following results.
\begin{theorem}
\label{theorem closed form solution}
If the optimization problem \textbf{(SEE3)} is feasible, the optimal power allocation policy maximizing the secrecy energy efficiency of the considered downlink multi-user NOMA system with single BD is obtained in closed-form as follows
\begin{align}
p_{k}^{*}(\alpha) &=(A_{k}-1)\left(\frac{1}{H_{k}^{(1)}(\rho^{*})} + p_{K}^{*}(\alpha) \prod\limits_{i=k+1}^{K-1}A_{i}\right. \nonumber \\
&\left. \qquad \qquad \ \ \  + \sum\limits_{i=k+1}^{K-1} \frac{A_{i}-1}{H_{i}^{(1)}(\rho^{*})}\prod\limits_{j=k+1}^{i-1}\!\!A_{j}\!\right)\!,   k \in \llbracket 1,K-1\rrbracket,
\nonumber \\
p_{K}^{*}(\alpha) &=\min\left( \max \left(\widehat{p}_{K}(\alpha),\frac{A_{K}-1}{H_{K}^{(1)}(\rho^{*})}\right),u\right) ,
\label{eq:optimal solution NOMA}
\end{align}
where $u \!\!=\!\!\frac{1}{\prod\limits_{i=1}^{K-1}A_{i}}\left(P_{\max} - P_{\min} + \frac{A_{K}-1}{H_{K}^{(1)}(\rho^{*})}\prod\limits_{j=1}^{K-1}A_{j}\right),$ and $\widehat{p}_{K}(\alpha)$  is the unique root of the derivative $\partial_{p_K}\Psi(p_{1}^{*},p_{2}^{*},\dots,p_{K})$ given in appendix B on~\cite{savard2024secrecy}. 
\end{theorem}
In conclusion, we have established:
(\romannumeral 1) an exact closed-form solution for the optimal backscatter coefficient $\rho^*$ (see ~\eqref{eq:optimal RC}); (\romannumeral 2) an analytical expression for the optimal power allocation (Theorem~\ref{theorem closed form solution}).
The special feature of our approach lies in its reduction to a one-dimensional problem via calculation of the critical point $\widehat{p}_{K}(\alpha)$, eliminating the need for iterative numerical optimization. Unlike exhaustive methods (of complexity $\mathcal{O}(2^K)$), our solution maintains a constant complexity $\mathcal{O}(1)$ whatever the number of users $K$, as the complexity analyses in Section~\ref{sec:res} will demonstrate.


\subsubsection{\textbf{Secrecy Energy-Efficiency as the Ratio Between Secrecy Sum Rate and Power Consumption}}
Our closed-form solution $\left(\rho^{*},\mathbf{p}^{*}\right)$ for \textbf{(SEE1)}  can also be used to optimize another metric for SEE, which is defined as the ratio of the achievable sum secrecy rate to the total power consumption as in~\cite{yao2019secrecy}:
\begin{equation}
\label{ratio secrecy energy efficiency} 
\zeta_{SEE}\left(\rho,\mathbf{p}\right) = \frac{\sum_{k=1}^{K}R_{k}^{s}\left(\rho,\mathbf{p}\right)}{\sum_{k=1}^K p_k + P_{c}}.
\end{equation} 

Since only the numerator, i.e., the secrecy sum-rate, depends on the RC \( \rho \), the optimal \( \rho^{*} \) that maximizes the objective in \textbf{(SEE3)} also maximizes \( \zeta_{SEE}(\rho, \mathbf{p}) \) for any given power allocation vector \( \mathbf{p} \).

Once the optimal RC \( \rho^* \) is determined, we can proceed to find the optimal power allocation that maximizes \( \zeta_{SEE}(\rho^{*}, \mathbf{p}) \). Given that the numerator of \( \zeta_{SEE}(\rho^{*}, \mathbf{p}) \) is a concave function of \( \mathbf{p} \), while the denominator is linear in \( \mathbf{p} \), we can apply fractional programming~\cite{isheden2012framework}. Specifically, maximizing \( \zeta_{SEE}(\rho^{*}, \mathbf{p}) \)  reduces to finding the solution to the following equation w.r.t. $\alpha$:
\begin{equation}\label{eq:F(alpha)}
F(\alpha) = \sum_{k=1}^{K} R_{k}^{s}\left(\rho^{*}, \mathbf{p}^{*}\right) - \alpha \left(\sum_{k=1}^{K} p_{k}^{*} + P_{c}\right) = 0,
\end{equation}
where \( \mathbf{p}^* \) is the closed-form solution to \textbf{(SEE3)}, as presented in Theorem~\ref{theorem closed form solution}. This optimization is efficiently solved using Dinkelbach's algorithm~\cite{zappone2015energy}.

In the following section, we present the closed-form solutions for the reflection coefficient and power allocation problem in the scenario involving two backscatter relays.

\subsection{Closed-Form Secrecy Energy-Efficiency Solution for Two-User NOMA with two Backscatter Devices}
For the case  involving two backscatter devices (BDs), although \textbf{(SEE2)} is not a convex problem, we present its closed-form analytical solution below. Our approach follows a similar methodology to the one used in~\cite{el2022energy}, where the energy efficiency is maximized for two users assisted by two BDs.
The key difference between our work
and that of~\cite{el2022energy} lies in the presence of a passive eavesdropper in our system model, which introduces
secrecy considerations and accordingly modifies the objective function.  
Following a similar strategy, we begin by decoupling the \textbf{(SEE2)} problem into two sub-problems:
 (\romannumeral 1) for a fixed power allocation vector $\mathbf{p} = (p_{1}, p_{2})$,  we determine the optimal reflection coefficients (RCs); (\romannumeral 2) using the resulting optimal RCs $\bm{\rho^{*}} = (\rho_{1}^*,\rho_{2}^*)$, we then optimize the power allocation vector $\mathbf{p}$. 
 
\subsubsection{\textbf{Optimal Reflection Coefficient}}
In the same spirit as in~\cite{el2022energy}, to optimize the RC vector, we have demonstrated in Appendix~\ref{appendix} the sign behavior of the objective function in \textbf{(SEE2)}. We distinguish two cases: in the first case, the objective function is monotonically increasing w.r.t. $\rho_{1}$ for fixed $p_{1}$, $p_{2}$, and $\rho_{2}$, and monotonically decreasing w.r.t. $\rho_{2}$. The second case is symmetric, where the objective function is monotonically increasing w.r.t. $\rho_{2}$ and monotonically decreasing w.r.t. $\rho_{1}$. This implies that the optimal RC solution, $\rho_{1}^{*}$ and $\rho_{2}^{*}$, lies on the Pareto boundary of the feasible set, as presented in the following theorem.
\begin{theorem}
\label{theorem closed form coefficient reflection solution}
Given a fixed power allocation vector $\mathbf{p} = (p_{1},p_{2})$, the optimal reflection coefficients $\rho_1^{*}$ and $\rho_2^{*}$ that maximize the secrecy energy-efficiency objective in \textbf{(SEE2)} are given by the following closed-form expressions:\\
(\romannumeral 1) If $
\left( \frac{G_{12}}{G_{1e}} > \frac{G_{22}}{G_{2e}} \!\right) 
\text{ and } 
  \left( \frac{G_{22}}{G_{2e}} \neq \frac{G_2}{G_e}\right) 
     \text{ and }\left( \frac{G_{12}}{G_{1e}} \neq \frac{G_2}{G_e} 
     \right)$, then four cases may arise, as detailed in Appendix~\ref{appendix}:\\ 
     \textbf{[H1]} If $(G_{11} - G_{12}) \leq 0$ and $(G_{21} - G_{22}) \leq 0$, then $\rho_1^* = 1$, $\rho_2^* = 0$.\\
     \textbf{[H2]} If $(G_{11} - G_{12}) > 0$ and $(G_{21} - G_{22}) \leq 0$, then   $\rho_1^* = \min\left(1, \rho_{1,\inf}\right)$, $\rho_2^* = 0$.\\
     \textbf{[H3]} If $(G_{11} - G_{12}) \leq 0$ and $(G_{21} - G_{22}) > 0$, then $\rho_1^* = 1$, $\rho_2^* = 0$.\\
     \textbf{[H4]} If $(G_{11} - G_{12}) > 0$ and $(G_{21} - G_{22}) > 0$, then $\rho_1^* = \min(1, \rho_{1,\inf})$ and $\rho_2^* = 0$.\\
     (\romannumeral 2) If $
\left( \frac{G_{12}}{G_{1e}} < \frac{G_{22}}{G_{2e}} \!\right) 
\text{ and } 
  \left( \frac{G_{22}}{G_{2e}} \neq \frac{G_2}{G_e}\right) 
     \text{ and }\left( \frac{G_{12}}{G_{1e}} \neq \frac{G_2}{G_e} 
     \right)$ the result follows by symmetry, with the roles of $\rho_1$ and $\rho_2$ interchanged.
Here, $\rho_{1,\inf}$ denotes the extremum value of $\rho_{1}$ when $\rho_{2}=0$ , whose explicit expression is provided in Appendix~\ref{appendix}.
\end{theorem}
\begin{IEEEproof}
The proof follows a similar methodology to that in~\cite{el2022energy}. The first step consists in analyzing the sign of the objective function in \textbf{(SEE2)} with respect to $\rho_1$ and $\rho_2$. Two distinct cases are identified, as detailed in Appendix~\ref{appendix}. Next, by applying a comparable reasoning to~\cite{el2022energy}, we show that the optimal RCs lie on the Pareto boundary of the feasible set.

The complete proof is provided in Appendix~\ref{appendix}. 
\end{IEEEproof}
\subsubsection{\textbf{Optimal Power Allocation}} 
Based on the closed-form expressions derived in Theorem~\ref{theorem closed form coefficient reflection solution}, and once the optimal RC vector $\bm{\rho}^{*} = (\rho_{1}^{*}, \rho_{2}^{*})$ is determined, we fix $\bm{\rho} = \bm{\rho}^{*}$ in the original problem \textbf{(SEE2)}. The problem then reduces to a power allocation sub-problem, formulated as follows
\label{SEE4}
\begin{align*}
\textbf{(SEE4)}~&  \max_{\mathbf{{p}}}   ~ \left( \sum_{k=1}^{2} R_{k}^{s}\left(\bm \rho, \mathbf{{p}}\right) - \alpha \left( \sum_{k=1}^{2} p_k + P_c \right) \right), \\
\text{ s.t. } 
&\theta_{1}(\mathbf{p})  \leq P_{\max} \tag{C1'} ,\\
   &\theta_{k} (\mathbf{p}) \geq A_k \theta_{k+1}(\mathbf{p}) + \frac{A_{k} - 1}{H_{k}^{(2)}(\bm{\rho^{*}})}, \forall k \in \{ 1, 2\}. \tag{C2'} 
\end{align*}
The closed-form solution to this power allocation problem follows the same structure as in the single backscatter scenario (see  \eqref{eq:optimal solution NOMA}), with the only modification being the replacement of the normalized  channel gain $H_k^{(1)}(\rho^*)$ by its dual-backscatter counterpart $H_k^{(2)}(\bm{\rho}^{*})$, as defined in~\eqref{normalizedchannelgain2BD}.

\begin{remark}
Since problem \textbf{(SEE4)} is convex and the closed-form solution $\mathbf{p}^*$ is obtained from the KKT conditions, which are both necessary and sufficient for optimality in convex optimization, we can conclude that this solution is globally optimal.
\end{remark}

In the following section, we extend our analysis to the general case of multi-user NOMA systems with multiple BDs, addressing the additional complexity and challenges introduced in this more realistic scenario.

\subsubsection{\textbf{General Case: Multi-User NOMA with Multiple BDs}}\label{subsubsec:GeneralCase}

In the general case of multiple users and multiple BDs, the optimization problem becomes considerably more complex. The  successive interference cancellation (SIC) decoding constraints impose a strict order among users based on their effective channel gains, which depend non-linearly on the reflection coefficients (RCs) of all the BDs. This strong interdependence makes it particularly challenging to analyze the feasibility of the decoding conditions for all users.

Even with a fixed number of BDs, increasing the number of users leads to a combinatorial growth in the number of SIC constraints, each depending on multiple coupled variables. Moreover, when the number of BDs exceeds two, it becomes impractical to visualize or analytically characterize the feasibility region of the RCs. This complexity renders conventional analytical or graphical methods insufficient for system-wide optimization.

To address this issue, we propose and compare two optimization approaches: an exhaustive grid search method (see Algorithm~\ref{alg:optimal_rho_SEE}) and a robust heuristic method based on the particle swarm optimization (PSO) algorithm~\cite{huang2025research,hamdi2025age} (see Algorithm~\ref{alg:pso_SEE_optimization}).


%
\begin{algorithm}[t]
\caption{Grid Search-Based Optimization of RCs for Multi-User Multi-Backscatter  \( \zeta_{SEE} \) Maximization}
\label{alg:optimal_rho_SEE}
\begin{algorithmic}[1]
    \Statex \textbf{Input:} $A$, $P_{\max}$, $h_{k}$, $h_{e}$, $g_{m}$, $g_{mk}$, $g_{me}$, $\forall k \in \{1,\ldots,K\}$ and $\forall m \in \{1,\ldots,M\}$, where $K$: number of legitimate users, $M$: number of backscatter devices.
    \Statex \textbf{Output:} Optimal reflection coefficient vector $\bm{\rho}^* = (\rho_{1}^*,\rho_{2}^*, \ldots, \rho_{M}^*)$.
    
    
    \State \textbf{Step 1: Coarse grid search (step size $\Delta_c = 0.1$) } 
    \State Define $\mathcal{R}_{\text{coarse}} = \{0, 0.1, 0.2, \ldots, 1\}$
    \State Initialize $\bm{\rho}_{\text{best}} \gets \mathbf{0}_M$, $\zeta_{\text{best}} \gets -\infty$
    \For {each combination $\bm{\rho}^{(c)} = (\rho_1^{(c)}, \ldots, \rho_M^{(c)})$ where $\rho_m^{(c)} \in \mathcal{R}_{\text{coarse}}$}
        \State Compute normalized channel $H_{k}^{(M)}(\bm{\rho}^{(c)})$ for all users
        \State Compute  power allocation  $\mathbf{p}(\bm{\rho}^{(c)})$ using eq.~\eqref{eq:optimal solution NOMA}
        \State Compute  $\zeta_{SEE}(\bm{\rho}^{(c)}  , \mathbf{p}(\bm{\rho}^{(c)}))$ using eq.~\eqref{ratio secrecy energy efficiency}  
        \If {$\zeta_{SEE}(\bm{\rho}^{(c)}, \mathbf{p}(\bm{\rho}^{(c)})) > \zeta_{\text{best}}$}
            \State $\bm{\rho}_{\text{best}} \gets \bm{\rho}^{(c)}$
            \State $\zeta_{\text{best}} \gets \zeta_{SEE}(\bm{\rho}^{(c)}, \mathbf{p}(\bm{\rho}^{(c)}))$         
        \EndIf
    \EndFor
    
    \State \textbf{Step 2: Fine grid search (step size $\Delta_f = 0.01$ around $\bm{\rho}_{\text{best}}$) }
    \State Initialize $\bm{\rho}^* \gets \bm{\rho}_{\text{best}}$, $\zeta^* \gets \zeta_{\text{best}}$
    \For{$m = 1$ \textbf{to} $M$}
        \State Define $\mathcal{R}_m^{\text{fine}} = \{\max(0, \rho_m^* - 0.1), \max(0, \rho_m^* - 0.1) + 0.01, \ldots, \min(1, \rho_m^* + 0.1)\}$
    \EndFor
    \For{each combination $\bm{\rho}^{(f)} = (\rho_1^{(f)}, \ldots, \rho_M^{(f)})$ where $\rho_m^{(f)} \in \mathcal{R}_m^{\text{fine}}$}
        \State Compute normalized channel $H_{k}^{(M)}(\!\bm{\rho}^{(f)}\!)$ for all users
        \State Compute  power allocation  $\mathbf{p}(\bm{\rho}^{(f)})$ using eq.~\eqref{eq:optimal solution NOMA}
        \State Compute  $\zeta_{SEE}(\bm{\rho}^{(f)}, \mathbf{p}(\bm{\rho}^{(f)}))$ using eq.~\eqref{ratio secrecy energy efficiency}
        \If {$\zeta_{SEE}(\bm{\rho}^{(f)}, \mathbf{p}(\bm{\rho}^{(f)})) > \zeta^*$}
            \State $\bm{\rho}^* \gets \bm{\rho}^{(f)}$
            \State $\zeta^* \gets \zeta_{SEE}(\bm{\rho}^{(f)}, \mathbf{p}(\bm{\rho}^{(f)}))$         
        \EndIf
    \EndFor
    
    \State \Return $\bm{\rho}^{*} = (\rho_{1}^*,\rho_{2}^*, \ldots, \rho_{M}^*)$, $\mathbf{p}^{*}(\bm{\rho}^{*})$, $\zeta_{SEE}\left(\bm{\rho}^{*},\mathbf{p}^{*}\right)$
\end{algorithmic}
\end{algorithm}
\begin{algorithm}[t]
\caption{
PSO-Based Optimization for Joint Reflection Coefficients and Power Allocation}
\label{alg:pso_SEE_optimization}
\begin{algorithmic}[1]
\Statex \textbf{Input:} $A$, $P_{\max}$, $h_{k}$, $h_{e}$, $g_{m}$, $g_{mk}$, $g_{me}$, $\forall k \in \{1,\ldots,K\}$ and $\forall m \in \{1,\ldots,M\}$, where $K$: number of legitimate users, $M$: number of backscatter devices;
\Statex \textbf{PSO Parameters: }swarm size $N_p$, maximum number of iterations $T_{\max}$, learning coefficients $c_1$ and $c_2$, inertia weights $w_{\min}$ and $w_{\max}$, maximum velocity $v_{\max}$, and penalty factor $\varsigma$.
\Statex \textbf{Output:} Optimal RC vector $\bm{\rho}^*$, power allocation vector $\mathbf{p}^*$, and the maximum secrecy energy-efficiency $\zeta^{*}_{SEE}$
\State \textbf{Step 1: Initialization}
 \For{each particle $i = 1$ to $N_p$}
        \State Initialize position $\bm{\rho}_i^{(0)} \sim \mathcal{U}[0,1]^M$
        \State Initialize velocity $\mathbf{v}_i^{(0)} \sim \mathcal{U}[-v_{\max},v_{\max}]^M$
        \State Compute power allocation $\mathbf{p}_i^{(0)}$ using eq.~\eqref{eq:optimal solution NOMA}
        \State Evaluate fitness $\zeta_i^{(0)} = \zeta_{SEE}(\bm{\rho}_i^{(0)}, \mathbf{p}_i^{(0)})$
        \State Set personal best $\mathbf{pbest}_i \gets \bm{\rho}_i^{(0)}$, $\zeta_{\mathbf{pbest},i} \gets \zeta_i^{(0)}$
    \EndFor
    \State Set global best: $\bm{\rho}^* \gets \arg\max_i \zeta_i^{(0)}$, $\zeta^{*}_{SEE} \gets \max_i \zeta_i^{(0)}$
    \State \textbf{Step 2: Iterative Optimization}
     \For{$t = 1$ to $T_{\max}$}
        \State Update inertia weight: \[w^{(t)} = w_{\max} - (w_{\max} - w_{\min}) \cdot \left(\frac{t}{T_{\max}}\right)^2\]
        \For{each particle $i = 1$ to $N_p$}
            \State Generate $r_1, r_2 \sim \mathcal{U}[0,1]^M$
            \State \textbf{Velocity update:}       
            $\mathbf{v}_i^{(t)} \!=\! w^{(t)} \cdot \mathbf{v}_i^{(t-1)} \!+ c_1 \cdot r_1 \cdot \text{\hspace{3cm}}(\mathbf{pbest}_i - \bm{\rho}_i^{(t-1)}) + c_2 \cdot r_2 \cdot (\bm{\rho}^* - \bm{\rho}_i^{(t-1)}\!)$
            \State Apply velocity bounds: \[\mathbf{v}_i^{(t)} \gets \min(\max(\mathbf{v}_i^{(t)}, -v_{\max}), v_{\max})\]

            \State \textbf{Position update:}
            $
            \bm{\rho}_i^{(t)} = \bm{\rho}_i^{(t-1)} + \mathbf{v}_i^{(t)}
            $
            \State Apply position constraints:\[\bm{\rho}_i^{(t)} \gets \min(\max(\bm{\rho}_i^{(t)}, 0), 1)\]

            \State Compute power allocation $\mathbf{p}_i^{(t)}$ using eq.~\eqref{eq:optimal solution NOMA}
            \State Evaluate fitness $\zeta_i^{(t)} = \zeta_{SEE}(\bm{\rho}_i^{(t)}, \mathbf{p}_i^{(t)})$

            \If{$\zeta_i^{(t)} > \zeta_{\mathbf{pbest},i}$}
                \State Update $\mathbf{pbest}_i \gets \bm{\rho}_i^{(t)}$, $\zeta_{\mathbf{pbest},i} \gets \zeta_i^{(t)}$
            \EndIf
            \If{$\zeta_i^{(t)} > \zeta^{*}_{SEE}$}
                \State Update $\bm{\rho}^* \gets \bm{\rho}_i^{(t)}$, $\mathbf{p}^* \gets \mathbf{p}_i^{(t)}$, $\zeta^{*}_{SEE} \gets \zeta_i^{(t)}$
            \EndIf
        \EndFor
    \EndFor\\
    \Return $\bm{\rho}^*$, $\mathbf{p}^*$, $\zeta^{*}_{SEE}$
\end{algorithmic}
\end{algorithm}

\subsubsection*{\textbf{Complexity Analysis}}
We now compare the computational complexity of the two optimization methods used in our study: (i) the exhaustive grid-based search and (ii) the PSO algorithm. Let \( M \) denote the number of backscatter devices, \( N_p \) the number of particles in PSO, and \( T_{\max} \) the maximum number of PSO iterations.

\paragraph{Grid-based search (Algorithm~\ref{alg:optimal_rho_SEE})}

This method performs a two-stage grid exploration:

\begin{itemize}
    \item \textbf{Coarse search:} Each reflection coefficient \( \rho_m \) is swept over the interval \([0,1]\) with a coarse step \( \Delta_c = 0.1 \), yielding \( \left( \frac{1}{\Delta_c} + 1 \right) = 11 \) possible values per device. Since all combinations are evaluated, the total number of coarse evaluations is $N_{\text{coarse}} = 11^M$. 
       For each of these combinations, the system computes the effective channels, determines the optimal power allocation using a closed-form expression, and evaluates $\zeta_{SEE}$. Assuming a constant cost $ C $ per evaluation, the complexity of the coarse search is $ \mathcal{O}(11^M \cdot C)$.
    
    \item \textbf{Fine search:} Around each coarse optimum \( \rho_m^* \), a refined search is performed within the interval \([ \rho_m^* - 0.1, \rho_m^* + 0.1 ]\) using a finer step \( \Delta_f = 0.01 \). This yields \( \left( \frac{0.2}{\Delta_f} + 1 \right) = 21 \) values per device and a total of $N_{\text{fine}} = 21^M $ combinations to evaluate, each with similar computational cost $ C $. Thus, the complexity of the fine search is $\mathcal{O}(21^M \cdot C)$.
\end{itemize}
\textbf{Total complexity:} The total computational cost of the exhaustive search is thus $\mathcal{O}\left( (11^M + 21^M) \cdot C\right)$. This exponential complexity in \( M \) renders the exhaustive grid search computationally prohibitive for large values of \( M \).

\paragraph{ Particle swarm optimization (Algorithm~\ref{alg:pso_SEE_optimization})}
The PSO algorithm maintains a swarm of \( N_p \) particles, each of which iteratively updates its position and velocity over \( T_{\max} \) iterations. At each iteration, each particle evaluates the fitness function $\zeta_{SEE}$ at its current candidate solution. Assuming each evaluation costs  $C$, the total number of evaluations is \( N_p \cdot T_{\max} \), and the overall complexity becomes:
$\mathcal{O}(N_p \cdot T_{\max} \cdot C)$.
This complexity is independent of the number of backscatter combinations and is thus significantly more scalable than the grid search, especially for large values of $ M $.
The complexity comparison of the two analysed algorithms is summarized in Table \ref{tab:complexity_summary}.
\begin{table}[t]
\renewcommand{\arraystretch}{1.6} 
\centering
\caption{Complexity Analysis of the Proposed Algorithms}
\begin{tabular}{|c|c|c|}
\hline
\textbf{Algorithm} & \textbf{Complexity} & \textbf{Growth w.r.t. }\( M \) \\
\hline
Grid-based search & \( \mathcal{O}\left( (11^M + 21^M) \cdot C \right) \) & Exponential \\
\hline
PSO algorithm & \( \mathcal{O}\left( N_p \cdot T_{\max} \cdot C \right) \) & Polynomial  \\
\hline
\end{tabular}
\label{tab:complexity_summary}
\end{table}
\section{XAI-assisted SEE Maximization}
\label{DeepLearning}
In this section, we propose a learning-based framework to both approximate and interpret the optimal resource allocation strategy that maximizes the  SEE in a downlink NOMA system with AmBC. 
To this end, we develop a two-stage methodology that combines supervised deep learning with XAI tools.

\subsection{FNN-based SEE Maximization}

In the context of our work, a two hidden layer FNN model~\cite{ref_dnn_comprehansive,alam2024secure} with $128$ neurons per layer is employed in order to efficiently optimize the power allocation and the reflection coefficient (RC) of the backscatter. The FNN inputs can be defined as $\ma{i} = [h_k, h_e, g_m, g_{mk}, g_{me}]^T$. In addition the FNN outputs depends on the desired application, and can be expressed as  $\ma{o}_{p} = [\hat{p}_{1}^{*},\hat{p}_{2}^{*},\dots,\hat{p}_{K}^{*}]^T$ and $\ma{o}_{\rho} = [\hat{\rho}_{1}^{*},\hat{\rho}_{2}^{*},\dots,\hat{\rho}_{M}^{*}]^T$,  corresponding to the power allocation and the RC estimation, respectively.
Finally, the mean squared error (MSE) is employed as a loss function. We note that the FNN model performs almost similarly to the optimal solution while reducing the latency and the overall computational complexity. 


While the proposed FNN can accurately approximate the optimal resource allocation policies, it is crucial to understand how the model makes its decisions, particularly in security-sensitive scenarios. To address this, we incorporate an explainability component based on SHAP, which is discussed in the following section.

\subsection{FNN Input Selection Using SHAP}

Several methods have been explored in the field of XAI to interpret the employed deep learning model. The objective is to assign a relevance score for
the input features and interpret how each feature affects the final decision. Hence, allowing end users and researchers to understand the specific outcomes provided by the model. SHAP is one of the feature selection XAI schemes that works according to
a game-theoretic permutation-based iterative approach to provide a global perspective by quantifying the
feature contributions across the whole dataset.

In our work, we employed SHAP scheme to assign relevance scores of the FNN model input features denoted by $x=\{ h_k, h_e, g_m, g_{mk}, g_{me} \}$, $\forall k \in \{1,\ldots,K\}$ and $\forall m \in \{1,\ldots,M\}$, where $K$ is the number of legitimate users and $M$ is the number of backscatter devices. SHAP introduces a unified approach for estimating Shapley values \cite{lundberg2017unified}, which fairly attributes the prediction to the features’ channel gains by considering all possible feature coalitions.
 In this context, input features (e.g., $h_k$, $h_e$, $g_m$, $g_{mk}$, $g_{me}$) are treated as players, and the FNN prediction function $\upsilon(x)$ is the characteristic function, which outputs the estimated optimal power allocation $[\hat{p}_{1}^{*},\hat{p}_{2}^{*},\dots,\hat{p}_{K}^{*}]$ and reflection coefficient $[\hat{\rho}_{1}^{*},\hat{\rho}_{2}^{*},\dots,\hat{\rho}_{M}^{*}]$. All possible coalitions of features are defined, and for each input feature, the marginal contribution to the prediction is computed after being added to every possible coalition of other features. 
 Thus, the Shapley value $\phi_j(v) $ of a feature $j$ is calculated by averaging its marginal contribution among all possible coalitions: 
\begin{equation}\label{EqSHAPval}
\phi_j(v) = \sum_{S \subseteq Q \setminus \{j\}} \frac{|S|! (L - |S| - 1)!}{L!} (v(S \cup \{j\}) - v(S)),
\end{equation}
where $L$ is the total number of input features (i.e., channel gains), $Q$ is the set of all features, $S$ is a subset of features excluding $j$, and $v(S \cup \{j\}) - v(S)$ is the marginal contribution of feature $j$ to the coalition $S$. The SHAP explanation model, denoted as $g(\boldsymbol{z}') $, is additive, meaning that
the prediction is expressed as the sum of the contributions of the individual features:
\begin{equation}
g(\boldsymbol{z}') = \phi_0 + \sum_{j=1}^L \phi_j z'_j,
\end{equation}
where $\boldsymbol{z}' = (z'_1, \ldots, z'_L)^T \in \{0,1\}^L$ is the coalition vector indicating the presence ($z'_j = 1$) or absence ( $z'_j = 0$) of feature $j$, $\phi_0$ is the expected output of the model, and $\phi_j$ is the Shapely value of feature $j$ as defined in {\eqref{EqSHAPval}}.

In our work, the instance to be explained is denoted by $x=\{h_{k}, h_{e}, g_{m}, g_{mk}, g_{me}\}$, where $L$ refers to the number of features included in $x$. Given the background dataset $D_{\text{SHAP}}$ and the employed black-box model $f$, 
the resulting SHAP procedure is summarized in Algorithm~\ref{alg:shap}. 

\begin{algorithm}[htbp]
\caption{SHAP Algorithm}
\label{alg:shap}
\begin{algorithmic}[1]
\State \textbf{Inputs}: $f$, $x$, $L$, $D_{\text{SHAP}}$
\State \textbf{Output}: SHAP values $\phi = (\phi_1, \ldots, \phi_L)$
\State Initialize an empty set of features to represent: $Z' = \emptyset$
\State Initialize target values for surrogate model: $Y = \emptyset$
\State Initialize weights for surrogate model: $W = \emptyset$
\For{$k = 1$ to $|D_{\text{SHAP}}|$}
    \State Sample a coalition $S$ 
    \State Sample a background instance $x_{\text{SHAP}}$ from $D_{\text{SHAP}}$
    \State Create masked instance $x_{\text{SHAP}}^{\prime}$:
    \State Get prediction from black-box model: $p_S = f(x_{\text{SHAP}}^{\prime})$
    \State Calculate Shapley weight for this coalition: $w_S = \frac{(L - |S| - 1)! |S|!}{L!}$
    \State Add $S$ to $Z'$
    \State Add $p_S$ to $Y$
    \State Add $w_S$ to $W$
\EndFor
\State Train a weighted linear regression model on $(Z', Y)$ with weights $W$:
    $$ g(z') = \phi_0 + \sum_{j=1}^L \phi_j z'_j $$
    Where $z'_j$ is 1 if feature $j$ is present, 0 otherwise.
\State \textbf{return} Coefficients $\phi_1, \ldots, \phi_L$ as SHAP values.
\end{algorithmic}
\end{algorithm}
\noindent
The additive explanation model in SHAP is given by
$$ f(x) \approx g(x') = \phi_0 + \sum_{j=1}^L \phi_j x'_j, $$
\noindent
where $g$ represents the surrogate model used to estimate the SHAP values. We note that SHAP offers a more robust and comprehensive explanation since it considers all feature interactions.
\section{Numerical Simulations}\label{sec:res} 
In this section, representative numerical results are presented to validate the theoretical analysis of 
our proposed solution, by comparing the secrecy energy-efficiency (SEE) defined as the ratio $\zeta_{SEE}$ across five scenarios: multi-user NOMA with $M = 1, 2, 3, \text{ and } 4$ backscatter devices (BDs), and multi-user OMA with $M = 2$ BDs. During simulations, users are randomly placed within circular area of radius $50$ meters centered at the source, while BDs are placed within a smaller circular area of radius $5$ meters. Given the limited range of ambient backscatter communication (AmBC) systems, we consider that communication links have a strong line-of-sight (LOS) component and fading-free channels characterized by a pathloss of the form  $d^{-\gamma}$~\cite{el2023multi,nazar2021ber,lin2024energy}, where $d$ represents the distance between two nodes and $\gamma$
 is the pathloss exponent. All presented results are averaged over $10^{4}$ different random configurations, each satisfying the feasibility conditions defined in Section \ref{feasiblecondition}. 
 The simulation
parameters are summarized in Table~\ref{tab:sim_params}.
Finally we note that our study covers the analysis of the following aspects: (\textit{i}) the impact of BDs on  $\zeta_{SEE}$, (\textit{ii}) the impact of imperfect CSI on the eavesdropper link, and (\textit{iii}) the impact of feature selection using the SHAP XAI method on the overall model performance.

\begin{table}[!t]
\renewcommand{\arraystretch}{1.1} 
\caption{Simulation Parameters}
\label{tab:sim_params}
\centering
\begin{tabular}{|c|c|}
\hline
\textbf{Parameter} & \textbf{Value} \\
\hline
\multicolumn{2}{|c|}{\textbf{System Parameters}} \\
\hline
Channel realizations & $10^{4}$  \\
\hline
Path-loss exponent ($\gamma$) & $3$ \\
\hline
Noise power ($\sigma_{k}^{2} = \sigma_{e}^{2}$) & $-30$ dBm \\
\hline
Minimum data rate ($R_{\min,k}=R_{\min}$) & $1$ bit/s/Hz \\
\hline
Total transmit power ($P_{\max}$) & $50$ dBm \\
\hline
Circuit power consumption ($P_{c}$) & $30$ dBm \\
\hline
\multicolumn{2}{|c|}{\textbf{PSO Parameters~\cite{wang2023joint}}} \\
\hline
Number of particles ($N_p$) & $30$ \\
\hline
Maximum iterations ($T_{\max}$) & $100$ \\
\hline
Inertia weight factor ($w_{\min}$) & $0.4$\\
\hline
Inertia weight factor ($w_{\max}$) & $0.9$\\
\hline
Learning factor $c_{1}, c_{2}$ & $1.5$\\
\hline
$v_{\max}$ & $\pi/8$\\
\hline
\end{tabular}
\end{table}
\subsection{Impact of Backscatter Devices on $ \zeta_{SEE}$} 
Note that,  unlike NOMA, under OMA the transmitter serves the $K$ users by dividing the time into equal slots, with each user assigned one~\cite{savard2024secrecy}. In this setting, the optimal reflection coefficient for the BD is known to be $\rho^{*}_{\text{OMA}} = 1$~\cite{el2023multi}. The corresponding optimal power allocation vector \(\mathbf{p}^{*,\text{OMA}}\) can be derived analytically, as discussed in Theorem 3 of~\cite{savard2024secrecy}. However, the explicit derivation is omitted here for conciseness.


\begin{figure}[!t]
    \centering
    \includegraphics[width=0.9\columnwidth, height=0.27\textheight]{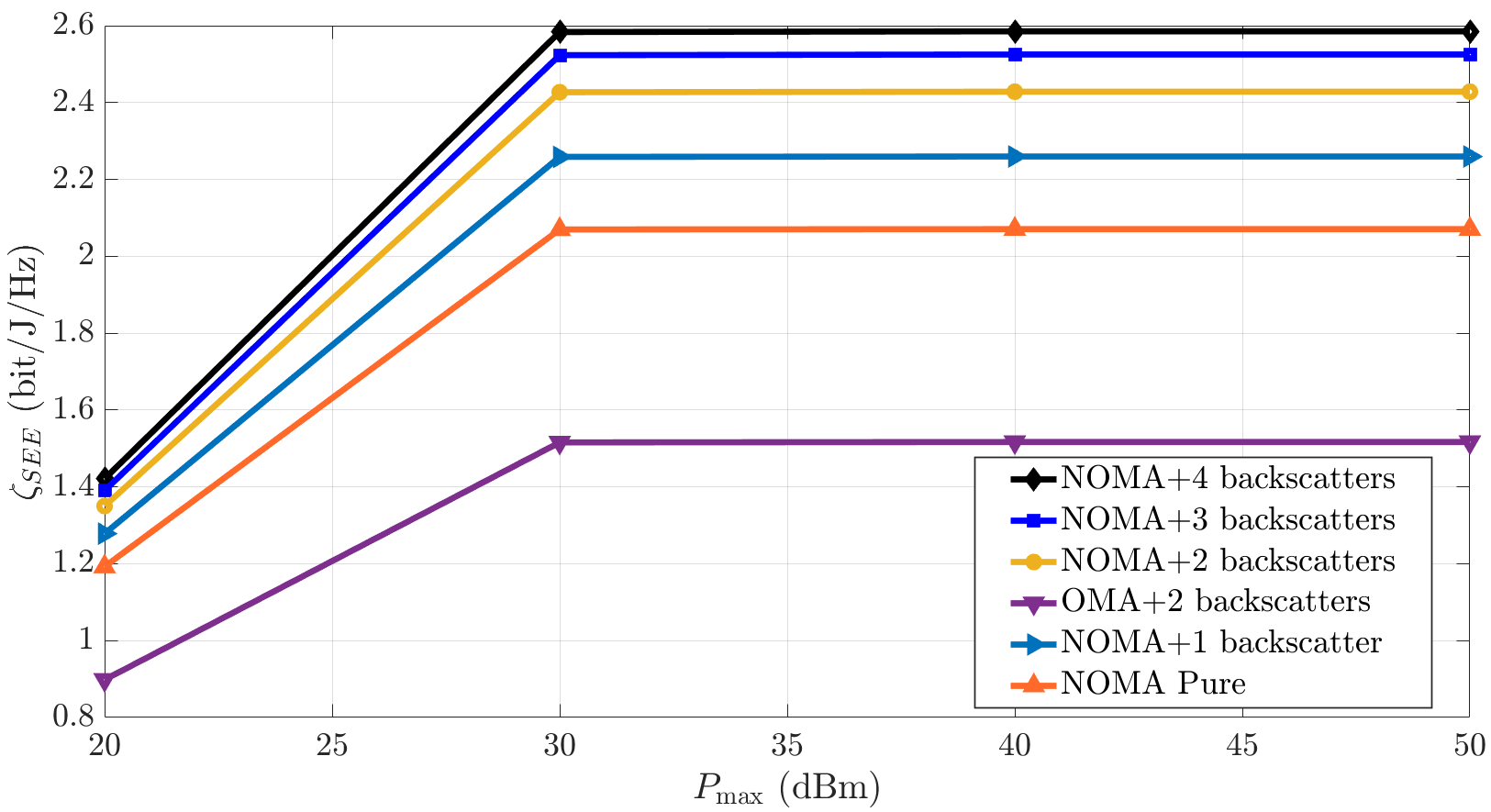}
    \caption{Secrecy energy-efficiency ratio $\zeta_{SEE}$ as a function of $P_{\max}$ for $K=2$ users and $R_{\min} = 1$ bit/s/Hz. An increase in the number of backscatter devices consistently improve the performance, regardless of $P_{\max}$.}
    \label{fig:SEE_Pmax}
\end{figure}
Fig.~\ref{fig:SEE_Pmax} depicts the evolution of the SEE ratio $\zeta_{SEE}$ as a function of the total transmit power $P_{\max}$, for $K=2$ users and a minimum rate constraint $R_{\min} = 1$ bit/s/Hz. The proposed scheme consistently outperforms all baseline approaches across the entire range of $P_{\max}$. Additionally, increasing the number of BDs leads to a significant improvement in $\zeta_{SEE}$ performance.
At low transmit power levels, the increase in secrecy sum-rate significantly contributes to the enhancement of $\zeta_{SEE}$. However, as $P_{\max}$ increases, the additional gains in secrecy rate become marginal, causing $\zeta_{SEE}$ to approach a saturation point.
\begin{figure}[!t]
    \centering
    \includegraphics[width=0.9\columnwidth, height=0.27\textheight]{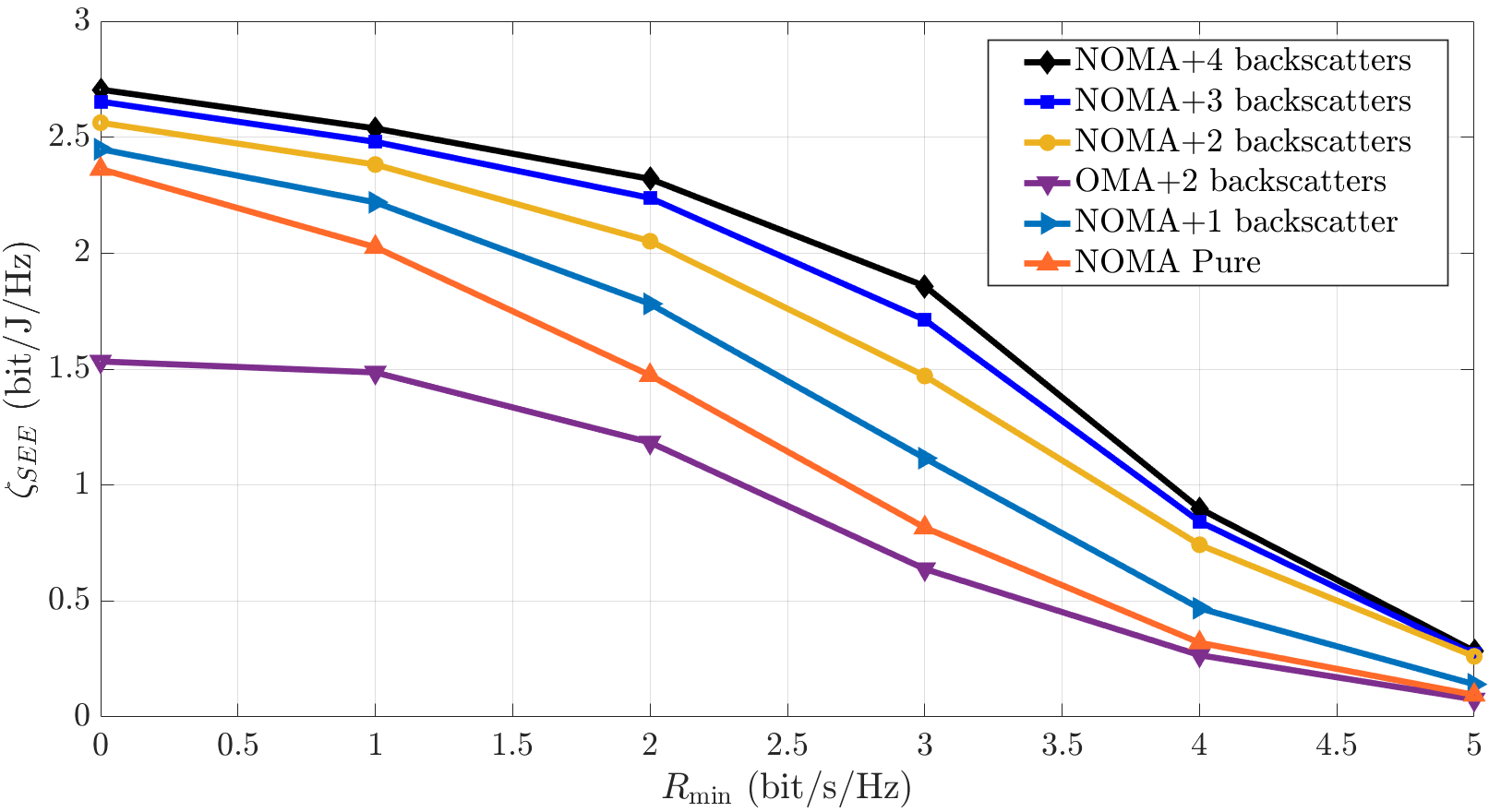}
    \caption{Secrecy energy-efficiency ratio $\zeta_{SEE}$ vs. $R_{\min}$ for $K=2$ users and $P_{\max} = 50$ dBm. The gain in $\zeta_{SEE}$ improves as the number of backscatter devices increases.}
    \label{fig:SEE_Rmin}
\end{figure}

Fig.~\ref{fig:SEE_Rmin} plots $\zeta_{SEE}$ versus the minimum QoS requirement $R_{\min}$ for $ K = 2 $ users and $P_{\max} = 50$ dBm. It is observed that the gain in $\zeta_{SEE}$ consistently improves with the increasing number of cooperative BDs. Moreover, for all schemes, $\zeta_{SEE}$ decreases as $R_{\min}$ increases. This is because increasing $R_{\min}$ requires the transmitter to allocate more power to enhance the data rate for users with poor channel conditions, which naturally degrades the $\zeta_{SEE}$ performance.
\begin{figure}[!t]
    \centering
    \includegraphics[width=0.9\columnwidth, height=0.27\textheight]{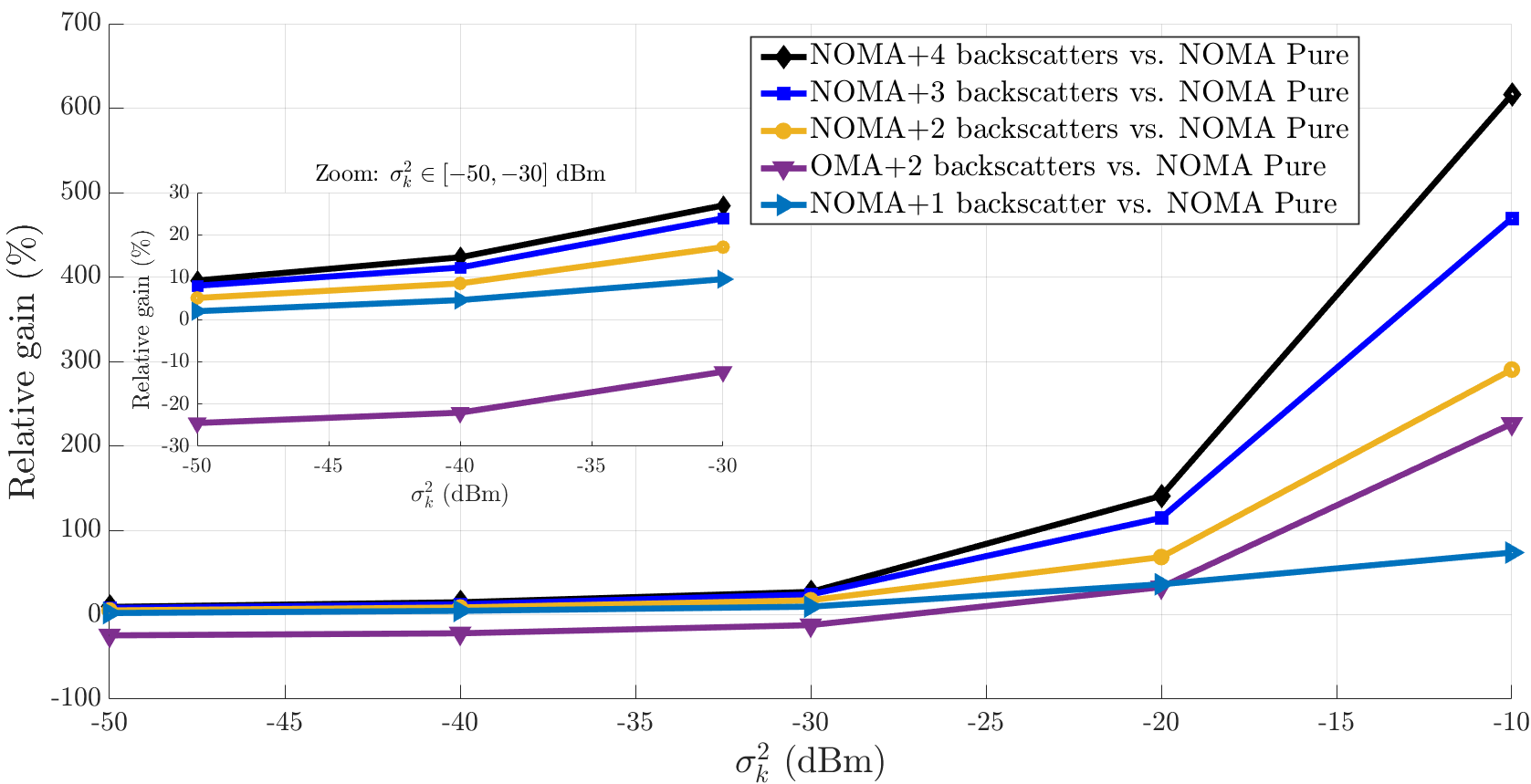}
    \caption{Relative secrecy energy-efficiency gain of NOMA with one to four BDs and OMA with two BDs, compared to pure NOMA, as a function of the noise variance \( \sigma_k^2 \), for \( K = 2 \) users, \( R_{\min} = 1 \) bit/s/Hz, and \( P_{\max} = 50 \) dBm. The gain increases with the number of BDs and becomes more significant in high-noise scenarios. For high values of \( \sigma_k^2 \), OMA with two BDs outperforms NOMA with a single BD.}
   \label{fig:RelativeGain_sigma}
\end{figure}

Fig.~\ref{fig:RelativeGain_sigma} illustrates the relative gain of various schemes compared to pure NOMA, where the relative gain is defined as  
$\big(\left(\zeta_{SEE}^{scheme} - \zeta_{SEE}^{NOMA}\right)/\zeta_{SEE}^{NOMA}\big)$ following~\cite{el2022energy}, and plotted as a function of the noise variance \(\sigma_{k}^{2}\). We see that, at the initial noise variance level of \(\sigma_{k}^{2} = -30~\text{dBm}\), the relative gains are approximately \(-16\%\), \(9\%\), \(17\%\), \(23\%\), and \(26\%\) for OMA with two BDs, NOMA with single BD, NOMA with two BDs, NOMA with three BDs, and NOMA with four BDs, respectively. As the noise variance increases beyond \(-20~\text{dBm}\), OMA with two BDs surpasses NOMA with a single BD in performance. Further increase in noise variance leads to a significant rise in relative gain, reaching up to \(615\%\) for NOMA with four BDs and \(470\%\) for NOMA with three BDs.
\begin{figure}[!t]
    \centering

    \begin{minipage}[b]{0.49\columnwidth}
        \centering
        \includegraphics[width=\linewidth]{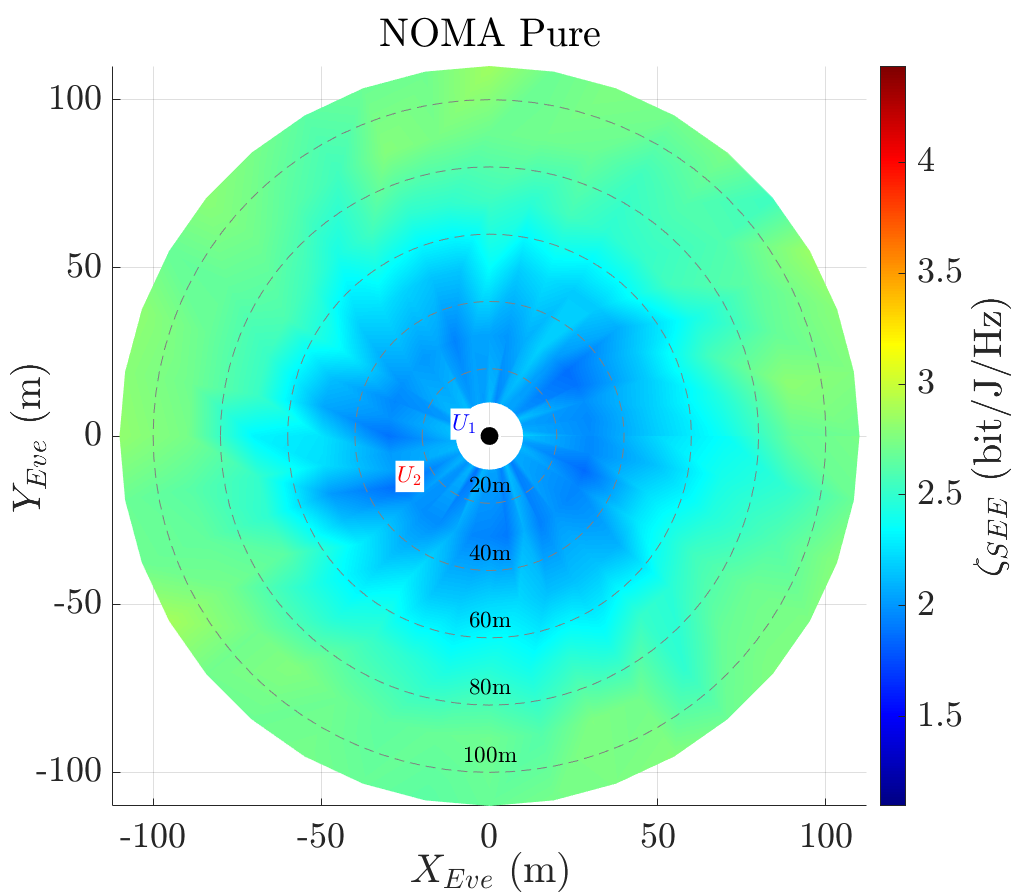}
    \end{minipage}
    \hfill
    \begin{minipage}[b]{0.49\columnwidth}
        \centering
        \includegraphics[width=\linewidth]{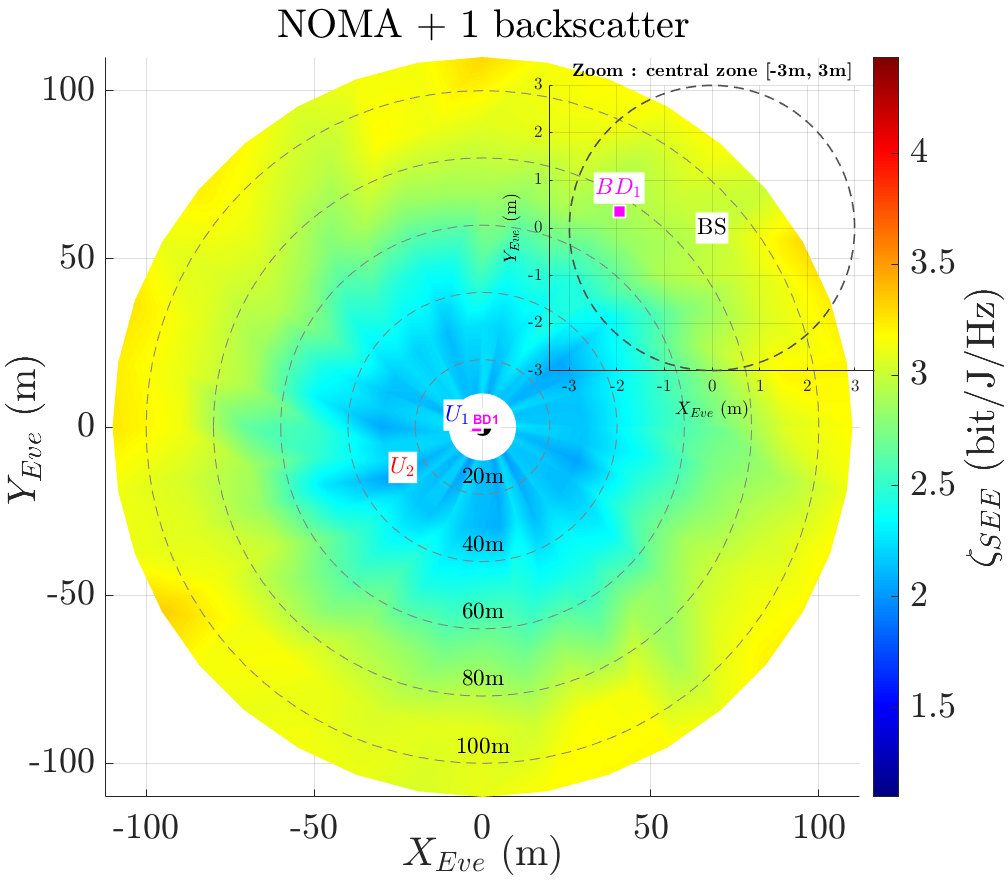}
        
    \end{minipage}
    
    \vspace{0.2cm}

    \begin{minipage}[b]{0.49\columnwidth}
        \centering
        \includegraphics[width=\linewidth]{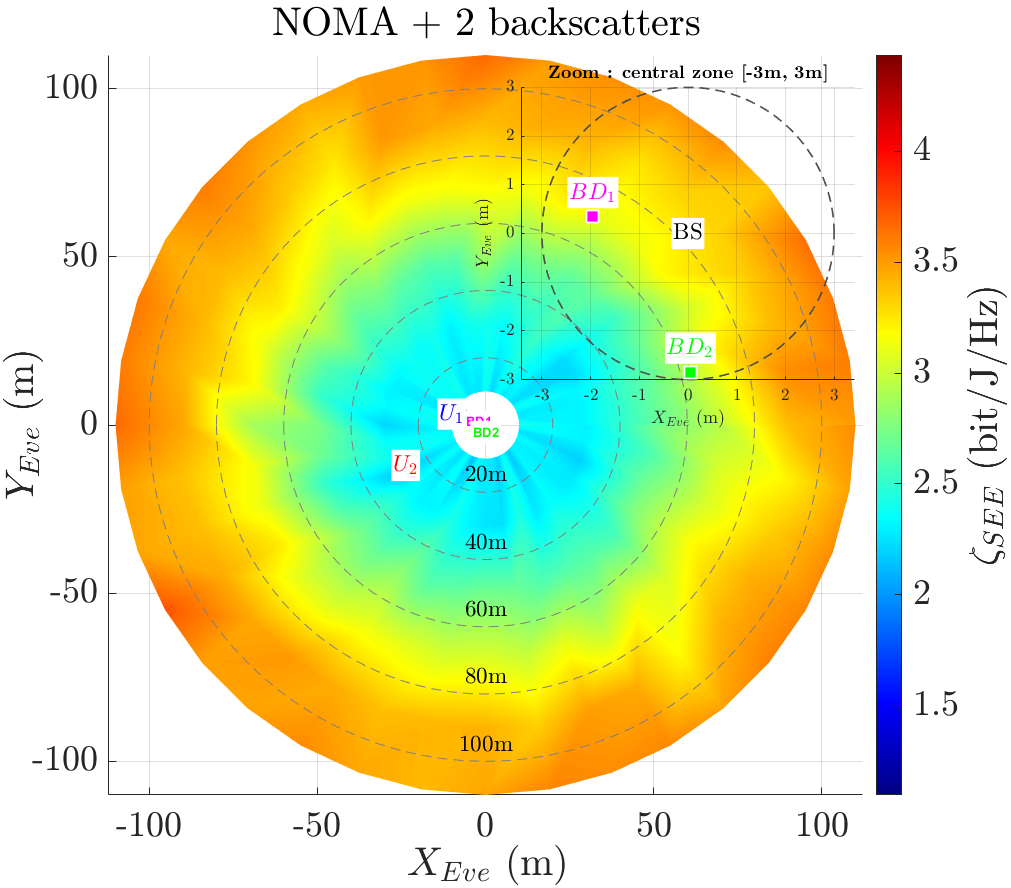}
        
    \end{minipage}
    \hfill
    \begin{minipage}[b]{0.49\columnwidth}
        \centering
        \includegraphics[width=\linewidth]{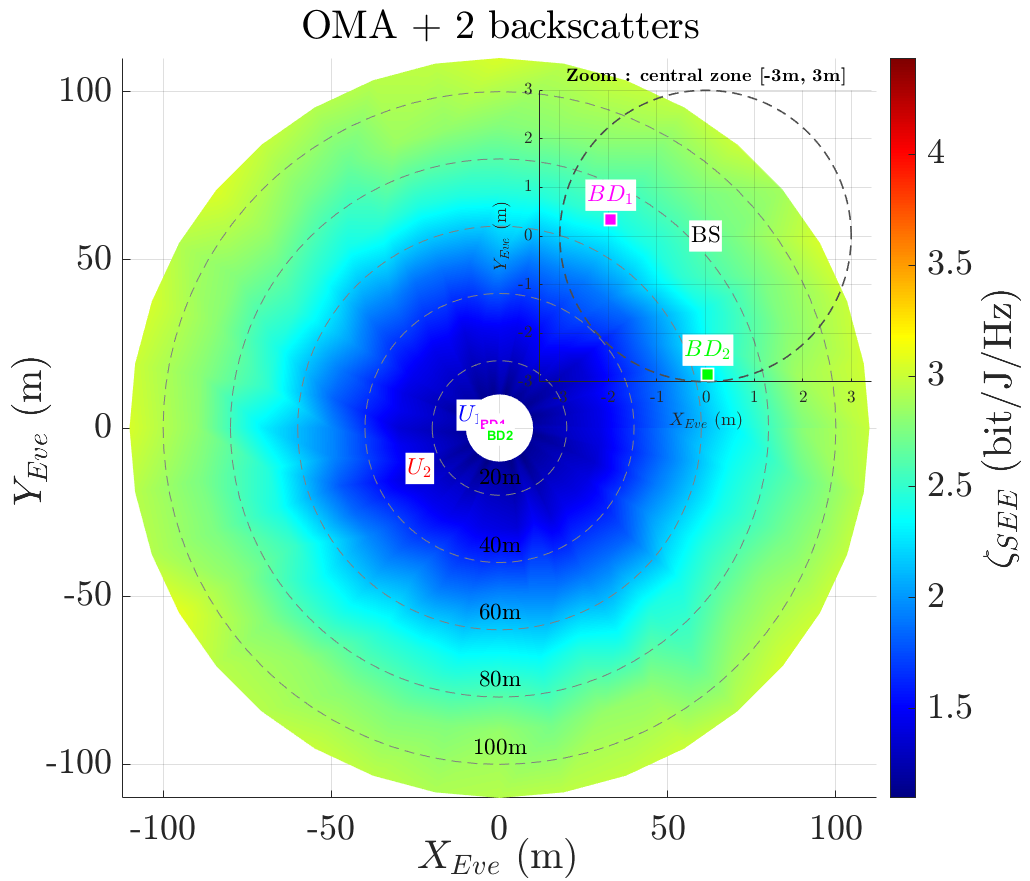}
      
    \end{minipage}

    \vspace{0.2cm}

    \begin{minipage}[b]{0.49\columnwidth}
        \centering
        \includegraphics[width=\linewidth]{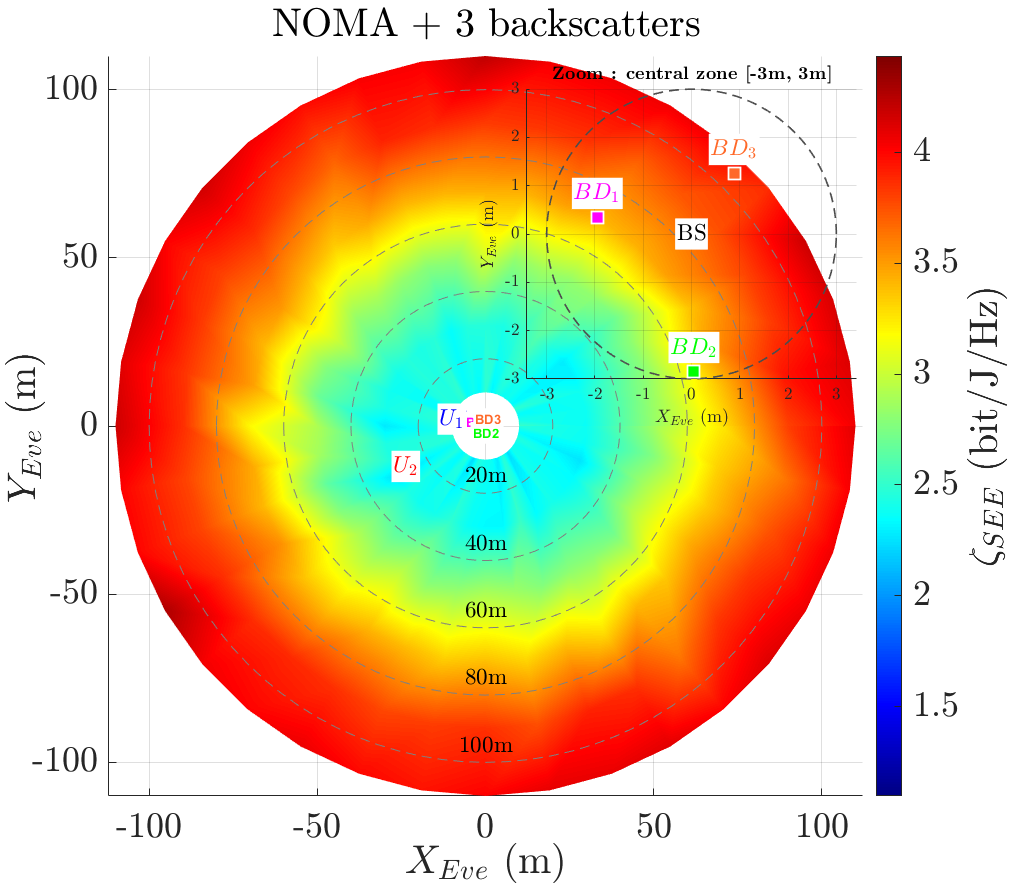}
        
    \end{minipage}
    \hfill
    \begin{minipage}[b]{0.49\columnwidth}
        \centering
        \includegraphics[width=\linewidth]{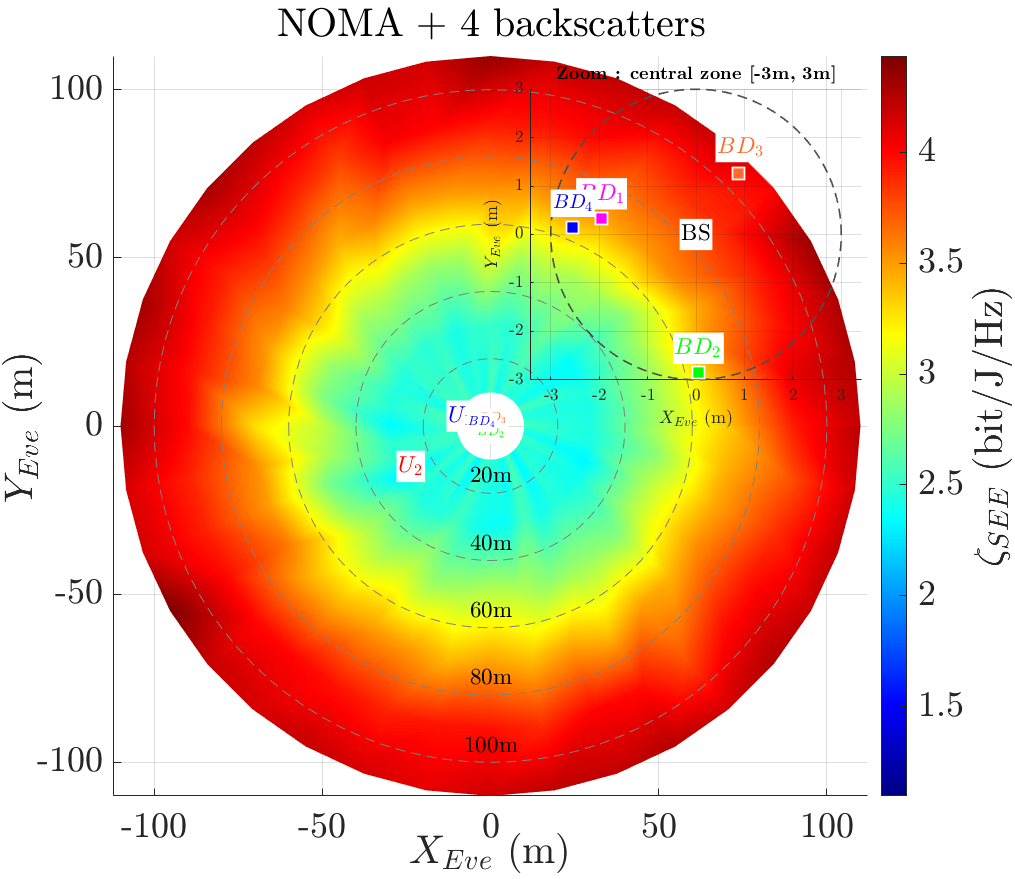}
        
    \end{minipage}
    \caption{Impact of the eavesdropper position on the secrecy energy-efficiency ratio $\zeta_{SEE}$. The integration of multiple BDs consistently improves $\zeta_{SEE}$ across all considered locations.}
    \label{fig:eavpos}
\end{figure}

Fig.~\ref{fig:eavpos} illustrates the variation of the SEE ratio $\zeta_{SEE}$ as a function of the eavesdropper's position in a two-user NOMA system assisted by a varying number of BDs. The results show that increasing the number of BDs consistently enhances the system's performance in terms of $\zeta_{SEE}$, regardless of the eavesdropper's location. With only one BD, the value of $\zeta_{SEE}$ is highly sensitive to the eavesdropper’s position, especially when the eavesdropper is located close to the transmitter or along strong propagation paths. However, when two or more BDs are deployed, the system becomes more robust, and the variations in $\zeta_{SEE}$ are significantly reduced. This stability is further improved as the number of BDs increases to three or four, highlighting the spatial diversity benefits offered by multiple backscatter paths. The observed improvement is mainly due to the joint optimization of the BDs' reflection coefficients, which enables the system to strengthen the legitimate users' received signals while simultaneously degrading the signal quality at the eavesdropper. Importantly, since BDs operate passively without active transmission, this gain in secrecy does not come at the cost of increased power consumption. It is also worth mentioning that having an
estimation of the distance to the eavesdropper could be sufficient to establish a communication strategy
that could reach a good SEE. With the increased sensing capabilities of networks in sixth generation (6G), tracking potential eavesdropper in specific areas could be feasible and allow to implement some of the allocation strategies we propose.

To conclude, all the presented results consistently demonstrate the advantages of integrating multiple BDs in improving the SEE of NOMA systems, regardless of system parameters such as transmit power, QoS constraints, noise variance, or the eavesdropper’s position. As the number of BDs increases, the system not only achieves higher secrecy performance but also becomes more resilient to adverse channel conditions and potential eavesdropping threats. However, it is also observed that the incremental gain between successive configurations tends to narrow as more BDs are added. This observation opens an important research question: how many BDs are truly optimal to balance performance enhancement with practical system complexity and deployment costs?

\subsection{Impact of Imperfect CSI on the Eavesdropper Link} 
In this part, we evaluate the robustness of our proposed scheme under imperfect channel state information (CSI) regarding the eavesdropper, focusing  specifically for the NOMA system aided by single BD. We assume that only an estimated channel $\hat{h}_e$ is available at the transmitter, while the actual eavesdropper's channel is modeled as $h_e = \hat{h}_e + \varepsilon$, where $\varepsilon \sim \mathcal{N}(0, \sigma_\varepsilon^2)$ represents the estimation error with zero mean and variance $\sigma_\varepsilon^2$. The system design including power allocation and backscatter reflection coefficients  is carried out using the estimated CSI $\hat{h}_e$, while the actual secrecy performance is evaluated based on the true channel realization $h_e$. This model reflects practical scenarios in which the transmitter has only
partial or imperfect knowledge of the eavesdropper's channel.
\begin{figure}[!t]
    \centering
   \includegraphics[width=0.9\columnwidth, height=0.27\textheight]{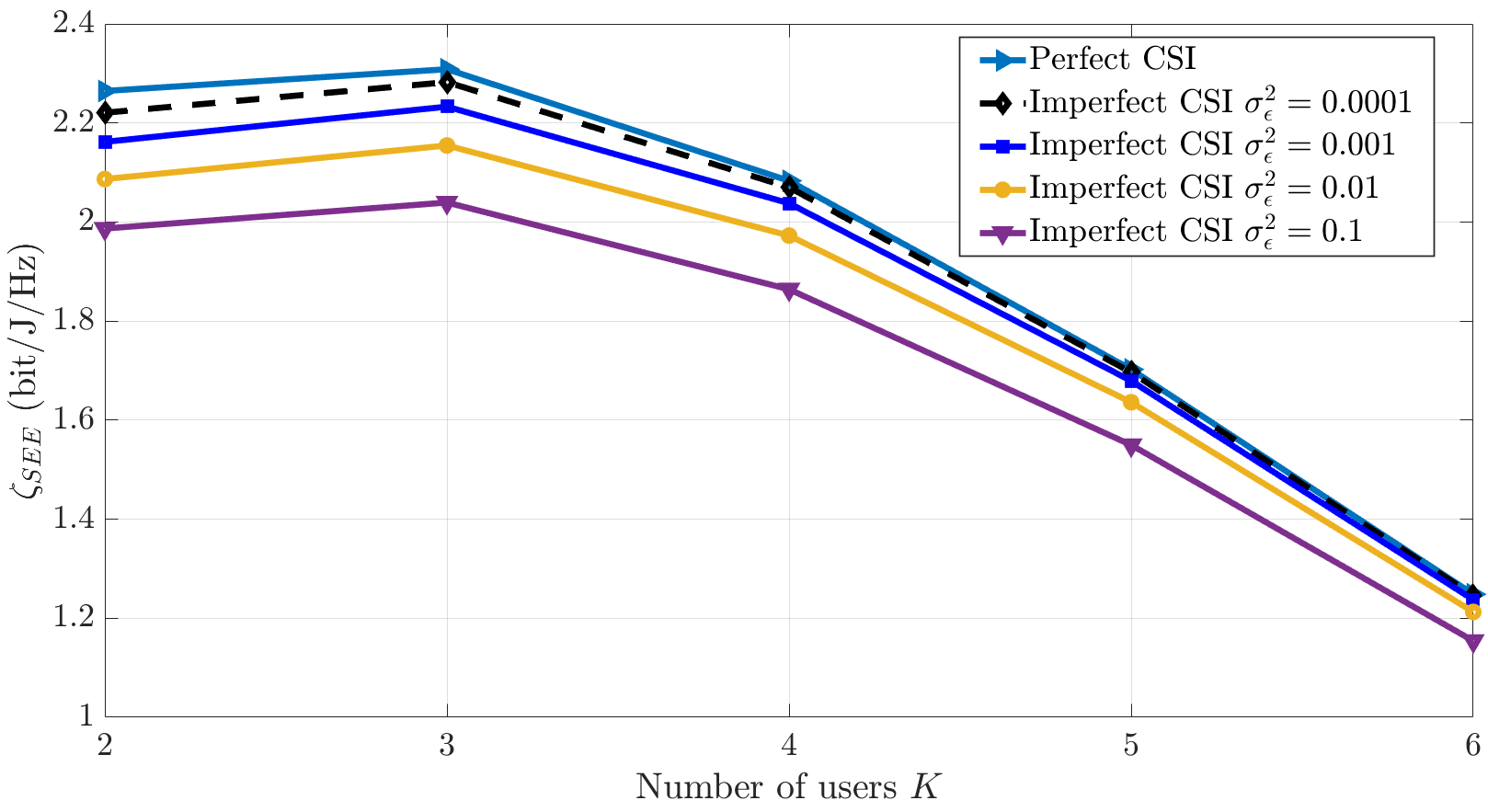}
    \caption{Impact of imperfect CSI on the secrecy energy-efficiency $\zeta_{{SEE}}$ for different values of estimation error variance $\sigma_\varepsilon^2$, as a function of the number of users $K$. The perfect CSI case is provided as a benchmark.}
    \label{fig:SEE_imperfectCSI}
\end{figure}

As illustrated in Fig.~\ref{fig:SEE_imperfectCSI}, the secrecy energy-efficiency $\zeta_{{SEE}}$ degrades progressively as the estimation error variance $\sigma_\varepsilon^2$ increases. This degradation becomes more significant as the number of users increases, due to the growing sensitivity of the power allocation strategy to inaccurate CSI. When $\sigma_\varepsilon^2$ is small (e.g., $\sigma_\varepsilon^2$ = $0.0001$), the performance remains close to the perfect CSI scenario, indicating a degree of robustness. However, as the estimation error becomes more significant (e.g., $\sigma_\varepsilon^2$ =  $0.1$), the $\zeta_{{SEE}}$ decreases substantially, highlighting the adverse effect of overestimating or underestimating the eavesdropper’s channel quality. This performance loss originates from suboptimal allocation decisions: either overly conservative (lower throughput) or overly aggressive (higher risk of information leakage), both leading to a reduction in energy efficiency. These results emphasize the importance of accurate CSI estimation, especially in systems with multiple users. To conclude, the proposed approach relies on a sufficiently accurate estimate of the eavesdropper’s channel. When such CSI is not reliably available, its uncertainty should be explicitly considered in both the problem formulation and the design of the optimization strategy.
\begin{figure*}[!t]
    \centering
    \begin{subfigure}[t]{0.48\textwidth}
        \centering
        \includegraphics[width=0.9\columnwidth, height=0.27\textheight]{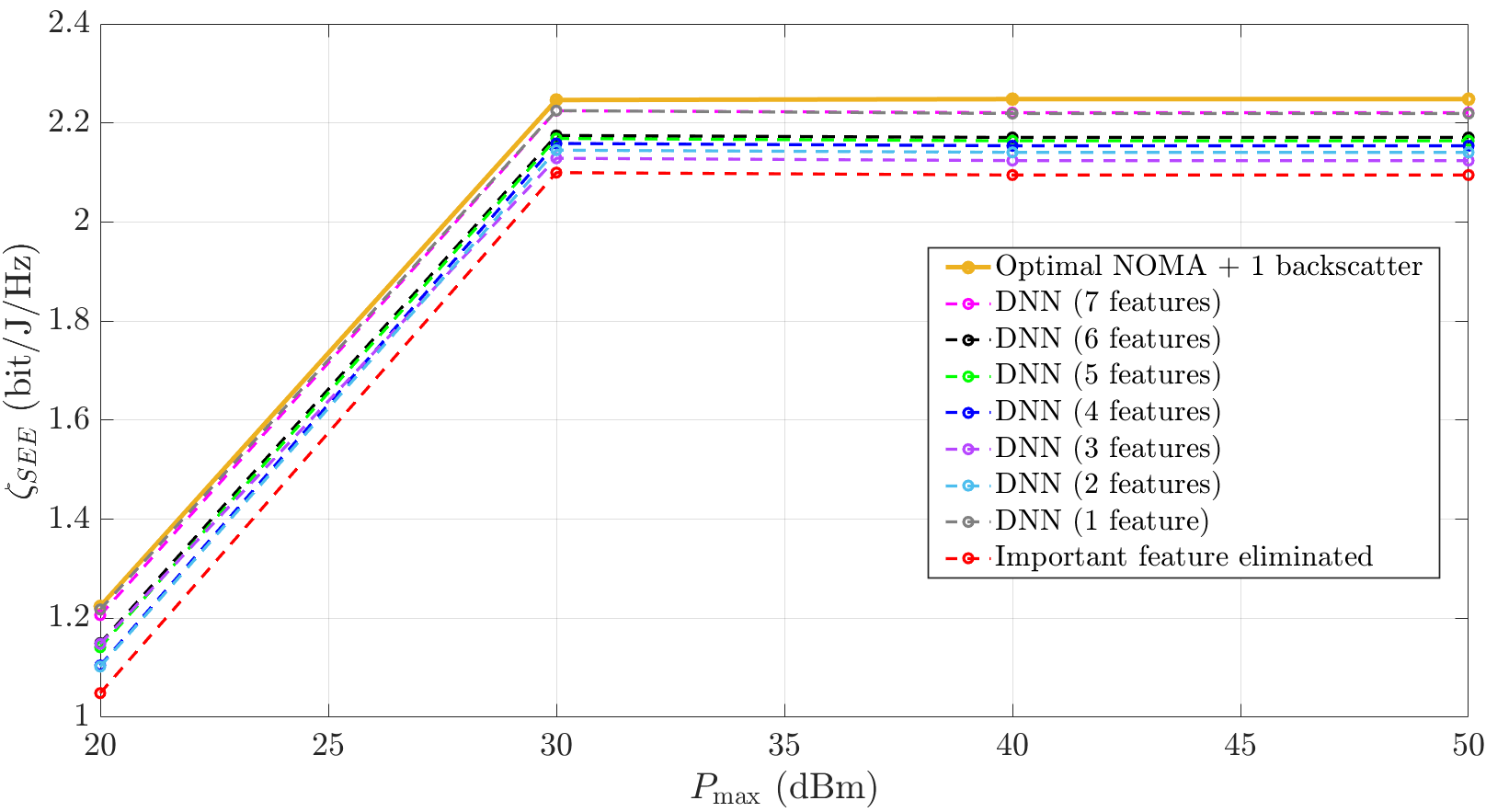}
        \caption{Impact of feature elimination on the predicted secrecy energy-efficiency ratio $\zeta_{SEE}$ as a function of $P_{\max}$ for $K=2$ users.}
        \label{fig:ablation_see}
    \end{subfigure}
    \hfill
    \begin{subfigure}[t]{0.48\textwidth}
        \centering
        \includegraphics[width=0.9\columnwidth, height=0.27\textheight]{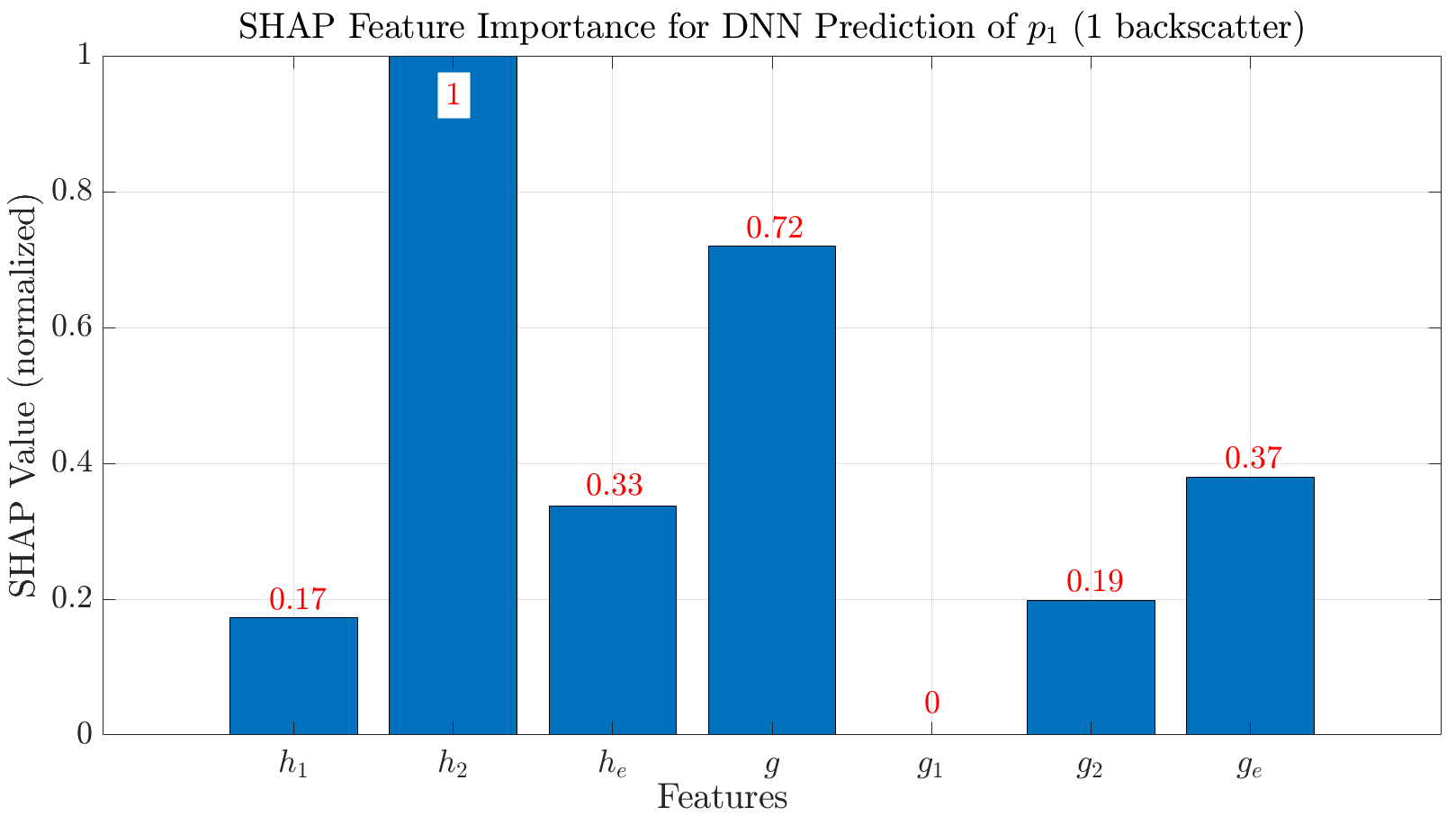}
        \caption{SHAP values for the prediction of transmit power $p_1$.}
        \label{fig:shap_p1}
    \end{subfigure}
    \caption{(a) Impact of SHAP-guided feature elimination on the predicted SEE ratio $\zeta_{SEE}$ as a function of $P_{\max}$ in a single backscatter-assisted NOMA system with $K=2$ users. (b) SHAP values indicating the contribution of each input feature to the prediction of transmit power $p_1$ under the same configuration.}
    \label{fig:ablation_and_shap}
\end{figure*}

\subsection{Model Performance and Explainability Analysis Using SHAP}

In this section, we employed a deep learning FNN model to jointly estimate the optimal reflection coefficients of the backscatter devices (BDs) as well as the power allocation. Moreover, to assess the physical relevance of the predictions, we used the SHAP XAI scheme to interpret the behavior of the FNN model by analyzing the relevant inputs that contribute efficiently to the desired prediction tasks.

\figurename~\ref{fig:ablation_see} shows the evolution of the SEE ratio $\zeta_{{SEE}}$ in two-user NOMA with single BD as a function of the maximum transmit power $P_{\max}$ for different FNN models built with a decreasing number of input characteristics. The ‘FNN ($7$ features)’ model uses the full set of seven channel gains, while the other models are obtained by successively removing the least influential variables based on the SHAP ranking provided in \figurename~\ref{fig:shap_p1}. In addition, a critical curve is included, corresponding to a FNN model trained without the single most important feature. Firstly, we observe that the FNN model using all the features closely approaches the optimal solution, thus validating its effectiveness. Furthermore, the gradual elimination of input features leads to a gradual degradation of $\zeta_{{SEE}}$, indicating that the network relies on all variables to deduce efficient allocation strategies. In particular, when the most important feature (as identified by SHAP in \figurename~\ref{fig:shap_p1}) is removed, a sharp drop in performance occurs, much greater than when less relevant features are omitted. This highlights the critical role played by certain input variables in enabling the model to capture the underlying structure of the system.


\figurename~\ref{fig:shap_p1} presents the normalized SHAP values associated with predicting the optimal transmit power $p_1$ in a two-user NOMA system assisted by a single BD. Each bar quantifies the average marginal contribution of an input characteristic (i.e. channel gain) to the model output. Interestingly, the direct channel gain for user~$2$ (i.e., $h_2$) appears to be one of the most influential features in predicting $p_1$, which may be attributed to its role in enforcing successive interference cancellation (SIC) constraints - a feature of NOMA systems.

Additionally, the channel gain between the transmitter and the BD (i.e., $g$) also exhibits a high SHAP score, which aligns with the structure of the composite channel gain $H_k^{(1)}$, where $g$ appears in a multiplicative form. Conversely, features such as the direct link to user~$1$ (i.e., $h_1$) and to the eavesdropper (i.e., $h_e$), although intuitively important, are ranked lower in SHAP importance. This can be explained by their indirect or compensatory influence, particularly when backscatter-assisted paths dominate the overall signal propagation.


\begin{figure*}[!t]
    \centering
    \begin{subfigure}[t]{0.48\textwidth}
        \centering
        \includegraphics[width=0.9\columnwidth, height=0.27\textheight]{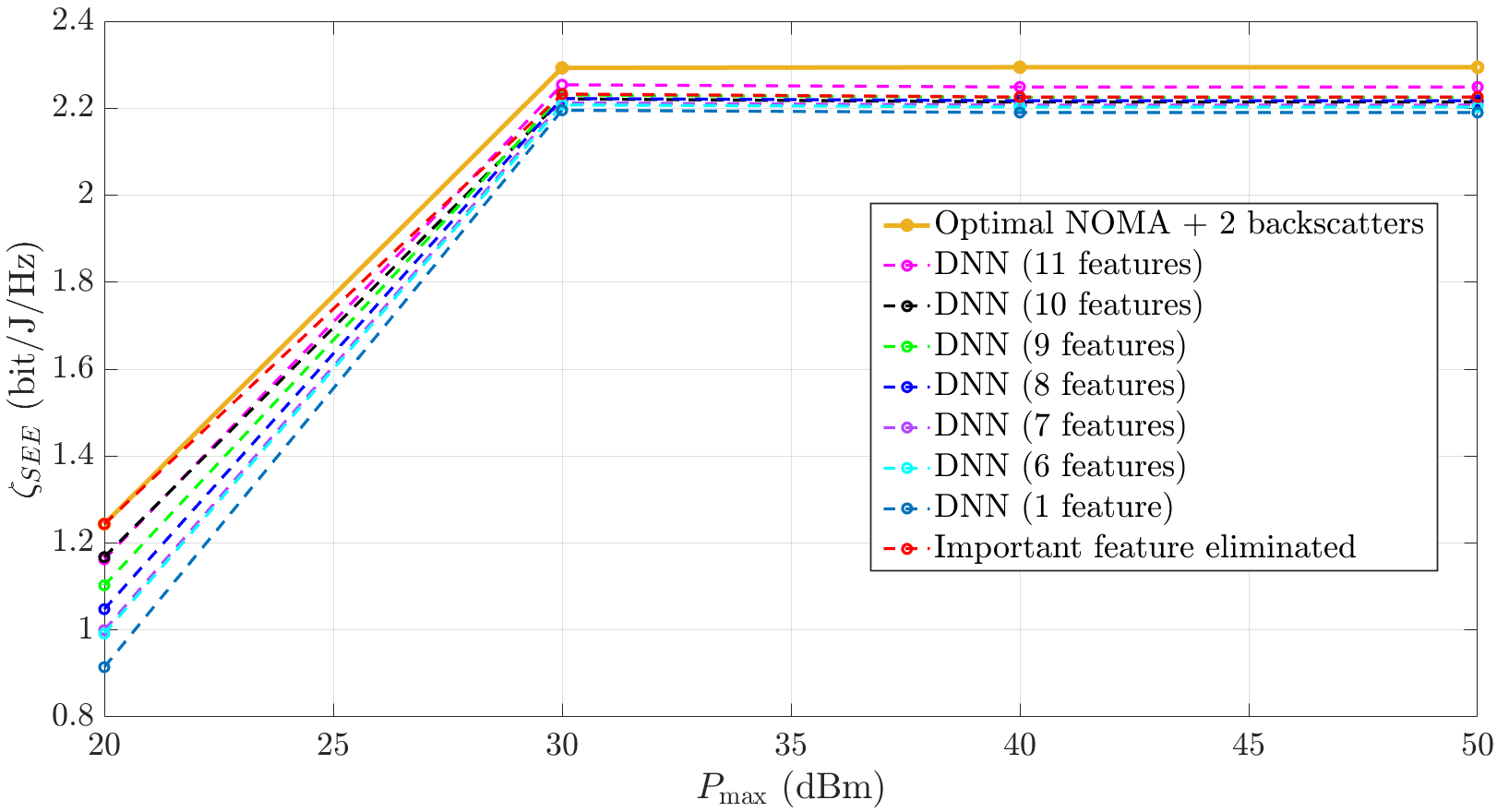}
        \caption{Impact of feature elimination on the predicted secrecy energy-efficiency ratio $\zeta_{SEE}$ as a function of $P_{\max}$ for $K=2$ users.}
        \label{fig:ablation2_see}
    \end{subfigure}
    \hfill
    \begin{subfigure}[t]{0.48\textwidth}
        \centering
        \includegraphics[width=0.9\columnwidth, height=0.27\textheight]{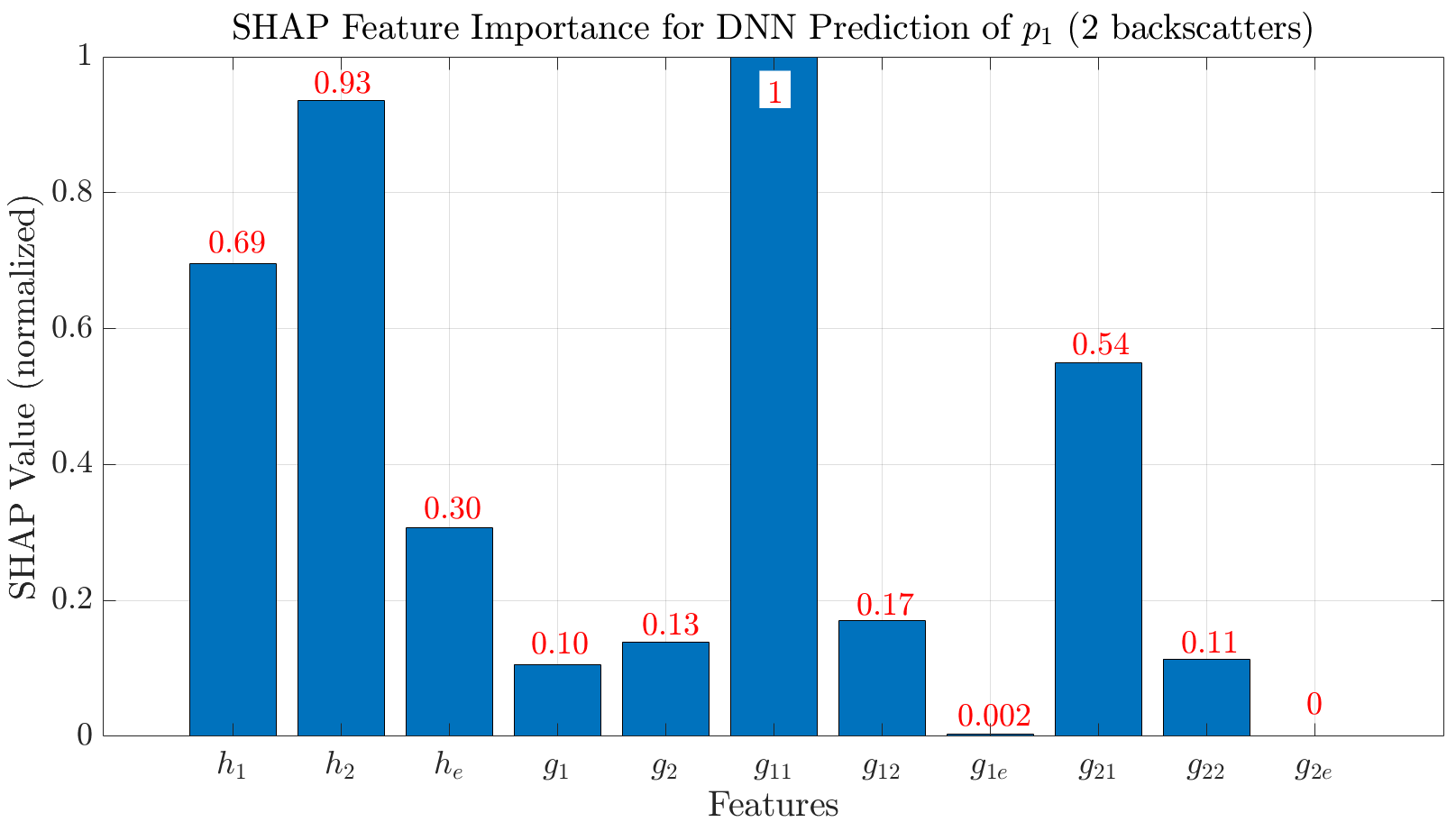}
        \caption{SHAP values for the prediction of transmit power $p_1$.}
        \label{fig:shap2_p1}
    \end{subfigure}
    \caption{(a) Impact of SHAP-guided feature elimination on the predicted SEE ratio $\zeta_{SEE}$ vs. $P_{\max}$ in a NOMA system assisted by two BDs with $K=2$ users. (b) SHAP values illustrating feature importance for predicting $p_1$ in this two-BDs scenario.}
    \label{fig:ablation_and_shap2}
\end{figure*}


Following the analysis of the single-BD scenario with seven input features, we now consider the two-BD configuration, in which the FNN is trained using all $11$ variables. 

\figurename~\ref{fig:ablation2_see} illustrates the effect of progressively removing inputs, ranked according to their SHAP importance as presented in \figurename~\ref{fig:shap2_p1}.
The ‘FNN ($11$ features)’ curve closely matches the ‘Optimal + $2$ backscatters’ benchmark, confirming that the model accurately approximates the optimal solution in this richer input space. 
Inputs are then eliminated progressively, starting with the least important, which leads to a decrease of $\zeta_{{SEE}}$ and highlights the cumulative contribution of multiple variables in determining efficient allocation policies. Notably, removing the most influential feature results in a performance drop of approximately $71\%$, substantially larger than the loss caused by excluding less relevant features. This observation highlights that, in the two-BD scenario, preserving this key feature is critical for maintaining SEE performance.

\figurename~\ref{fig:shap2_p1} shows the normalized SHAP values for predicting the optimal transmit power $p_{1}$. 
%
It is interesting to note that the channel gain between  $BD_{1}$ and user $1$, denoted by $g_{11}$, is the most influential feature. This result can be explained by the structure of the composite channel $H_{1}^{(2)}(\bm{\rho})$ defined in~\eqref{normalizedchannelgain2BD}, which models the effective channel at user $1$ as a non-linear combination of the direct channel $h_{1}$ and the
two backscattered components via $BD_{1}$ and $BD_{2}$. Since $g_{11}$ appears multiplicatively in this expression, it has a direct impact on the SINR of user $1$ and therefore plays a critical role in meeting the SIC constraints.  

The second most important feature is the direct link $h_{2}$ to user $2$. In the NOMA setting, this user typically benefits from the strongest channel and must be able to decode and cancel user $1$’s signal. As such, the feasibility of SIC heavily relies on the strength of $h_{2}$, justifying its high SHAP value. Next in importance are the backscatter-related links such as $g_{21}$, representing the link between $BD_{2}$ and user $1$. While less influential than $g_{11}$, these components enhance the channel diversity for user $1$ and can compensate for variations in the direct path, thus playing a meaningful role in the model’s power allocation decisions. 

In contrast, features corresponding to the eavesdropper's channels (e.g., $h_{e}$, $g_{1e}$, $g_{2e}$) exhibit relatively low SHAP values. Although these might seem intuitively important from a security standpoint, the system is designed to ensure secrecy even in worst-case conditions, primarily by reinforcing the legitimate links. Consequently, the model learns to maximize the secrecy rate by optimizing favorable channels, while the eavesdropper-related channels are handled implicitly through the overall system structure.
Lastly, gains such as $g_{12}$ and $g_{22}$, corresponding to the backscatter paths toward user $2$, show moderate influence. Their contribution is mostly tied to their effect on the SIC decoding at user $2$, yet it remains less critical than the dominant direct or composite channels.

Overall, the SHAP analysis aligns well with the theoretical formulation of the NOMA system assisted by dual backscatter. It highlights the key links responsible for effective SIC and secure communication, thereby reinforcing the physical consistency of the model and increasing confidence in its near-optimal decision-making capabilities. Moreover, further computational complexity reduction can be achieved by employing only the highlighted importance inputs by SHAP instead of the full input set. This issue depends mainly on the desired performance-complexity trade-off.

\section{Conclusion and Future Work}\label{Conclusion}
In this paper, we investigated the secrecy energy-efficiency (SEE) maximization problem in a downlink NOMA system assisted by ambient backscatter devices (BDs) in the presence of a passive eavesdropper. We first analyzed the configuration with a single BD, for which closed-form optimal solutions were derived for both the reflection coefficient (RC) and the power allocation. The study was then extended to the case of two BDs, where we showed that the optimal RCs lie on the Pareto boundary of the feasible region, enabling closed-form expressions for both the RCs and the power allocation policy. We further generalized the analysis to scenarios with an arbitrary number of BDs (greater than two). In this more complex setting, the optimization problem becomes analytically intractable due to the nonlinear coupling among variables and the constraints imposed by SIC decoding. To address these challenges, we proposed  two solution strategies: an exhaustive grid search method and a scalable PSO algorithm. Our results show that while the exhaustive search is suitable for small-scale systems, the PSO-based approach significantly reduces computational complexity and remains effective as the number of devices increases. To further enhance scalability and reduce latency, we developed a deep learning framework based on feedforward neural network (FNN) that accurately predicts the optimal RCs and power allocation. Moreover, we incorporated SHAP-based interpretability analysis, which revealed that the most influential input features align with the theoretical system model, thereby reinforcing trust in the proposed AI-driven solutions. Our numerical results show that  the inclusion of AmBC
significantly improves SEE, achieving gains of up to $615\%$ compared to conventional NOMA in high-noise regimes. Additionally, the FNN model achieves over $95\%$ accuracy compared to the optimal baseline while reducing computational complexity. As future work, it would be of great interest to extend the proposed framework to more practical AmBC-NOMA scenarios with imperfect CSI and multi-antenna architectures, where the joint design of beamforming, power allocation, RCs, and adaptive user clustering could further enhance SEE. Moreover, advanced AI models beyond feedforward networks, such as reinforcement learning, may be explored to ensure scalability, robustness, and real-time adaptability in large-scale networks.

\begin{appendices}
\numberwithin{equation}{section}
\section{Proof of Theorem~\ref{theorem closed form coefficient reflection solution} (Optimal Reflection Coefficient)}
\label{appendix}
We start by introducing the expression of the secrecy sum-rate ($SSR$) involved in the secrecy energy-efficiency optimization problem \textbf{(SEE2)}:
\begin{equation} 
    SSR(\bm \rho, \bm p ) \!=\!\! \sum_{k=1}^{2} R_{k}^{s}\left(\bm \rho, \mathbf{{p}}\right) \!=\! \frac{1}{2} \log_{2} \left( \frac{1 + H_{2}^{(2)}(\bm\rho) p_{2}}{1 + H_{e}^{(2)}(\bm\rho) p_{2}} \right),
\end{equation} 
where $H_{k}^{(2)}(\bm\rho)$ is defined in~\eqref{normalizedchannelgain2BD}. 

It is evident that only the $SSR(\bm\rho, \mathbf{p} )$ term in \textbf{(SEE2)} depends on the reflection coefficient vector $\bm \rho = (\rho_{1}, \rho_{2})$. Therefore, for a fixed power allocation $\mathbf{p}$, the optimization problem reduces to maximizing $SSR(\bm \rho, \mathbf{p})$ with respect to $\bm \rho$.

To this end, we analyze the behavior of the $SSR$ with respect to each component $\rho_m$ of the reflection vector, where $m \in \{1, 2\}$, while keeping the other coefficient $\rho_n$ ($n \neq m$) fixed.

The first-order derivative of $SSR$ with respect to $\rho_m$ is given by
\small
\begin{align}\label{firstderivativef1}
&\frac{\partial SSR}{\partial \rho_m} = 
\frac{1}{2\ln 2} \Bigg[ \frac{p_{2}\Big(\frac{\partial H_{2}^{(2)}(\rho_1,\rho_2)}{{\partial \rho_m}} - \frac{\partial H_{e}^{(2)}(\rho_1,\rho_2)}{\partial \rho_m}\Big)}{\Big(1 +H_{2}^{(2)}(\rho_1,\rho_2) p_{2}\Big)\Big(1 +H_{e}^{(2)}(\rho_1,\rho_2) p_{2}\Big)} \nonumber\\
& +\frac{ p_{2}^{2}\!\left(\!\frac{\partial H_{2}^{(2)}\!(\rho_1,\rho_2)}{\partial \rho_m}H_{e}^{(2)}\!(\rho_1,\rho_2)\!-\! \frac{\partial H_{e}^{(2)}\!(\rho_1,\rho_2)}{\partial \rho_m}H_{2}^{(2)}\!(\rho_1,\rho_2)\!\!\right)}
{\Big(1 +H_{2}^{(2)}(\rho_1,\rho_2) p_{2}\Big)\Big(1 +H_{e}^{(2)}(\rho_1,\rho_2) p_{2}\Big)}\!\Bigg],
\end{align}
\normalsize
where the terms $\frac{\partial H_{2}^{(2)}(\rho_1,\rho_2)}{\partial \rho_m}$ and $\frac{\partial H_{e}^{(2)}(\rho_1,\rho_2)}{\partial \rho_m}$ denote the partial derivatives of the normalized channel gains defined in~\eqref{normalizedchannelgain2BD} with respect to $\rho_m$.
These derivatives can be computed via straightforward differentiation and are omitted here for brevity. 
However, using standard properties of derivative signs, we can qualitatively determine the behavior of the $SSR$ function with respect to each reflection coefficient. In particular, we consider the following case:\\
\textbf{(\romannumeral 1)} If \small$
\left( \frac{G_{12}}{G_{1e}} > \frac{G_{22}}{G_{2e}} \!\right) 
\text{ and } 
  \left( \frac{G_{22}}{G_{2e}} \neq \frac{G_2}{G_e}\right) 
     \text{ and }\left( \frac{G_{12}}{G_{1e}} \neq \frac{G_2}{G_e} 
     \right)$,
\normalsize it can be shown that the secrecy sum-rate $SSR(\bm \rho, \mathbf{p} )$ is monotonically increasing w.r.t. $\rho_{1}$ for fixed $p_{1}$, $p_{2}$ and $\rho_{2}$ and monotonically decreasing w.r.t. $\rho_{2}$. 
Hence, the optimal reflection coefficients $(\rho_1, \rho_2)$ must lie on the Pareto boundary of the feasible set,  defined as the set of points where any further increase in $\rho_1$ or decrease in $\rho_2$ would violate at least one of the constraints (C2)–(C4) of \textbf{(SEE2)}.



From equation~\eqref{normalizedchannelgain2BD}, we observe that $H_k^{(2)}(\bm \rho)$ increases with both $\rho_1$ and $\rho_2$. Therefore, the right-hand side of constraint (C2) decreases as $\rho_1$ or $\rho_2$ increases, thus making this constraint easier to satisfy. 

Since the objective function is monotonically increasing w.r.t. $\rho_1$, maximizing $\rho_1$ improves the performance without compromising the feasibility of constraint (C2). In contrast, while reducing $\rho_2$ tends to improve the objective, it may also tighten constraint (C2), potentially leading to infeasibility. Therefore, the optimal value of $\rho_2$ must result from a careful trade-off between improving the objective and maintaining feasibility. Consequently, the maximization of $\rho_1$ can be addressed primarily through constraints (C3) and (C4), whereas the selection of $\rho_2$ must also account for constraint (C2).

To this end, and after some mathematical derivations, constraint (C3) can be reformulated as
\begin{equation}\label{Paretoinequality}
    \sqrt{\rho_1} (G_{11} - G_{12}) + \sqrt{\rho_2} (G_{21} - G_{22}) \leq G_2 - G_1.
\end{equation}

To better understand the feasible region and identify the optimal pair $(\rho_1, \rho_2)$, we focus on the equality corresponding to the boundary of this constraint:
\begin{equation}\label{eq:ParetoBoundary}
    \sqrt{\rho_1} (G_{11} - G_{12}) + \sqrt{\rho_2} (G_{21} - G_{22}) = G_2 - G_1.
\end{equation}

This boundary represents the Pareto frontier of the feasible set within the unit square $[0, 1]^2$. We adopt a geometric analysis of this frontier by examining how its shape and position depend on the signs of the coefficients $(G_{11} - G_{12})$ and $(G_{21} - G_{22})$. Four distinct configurations arise based on the signs (positive, negative, or zero) of these terms, and each configuration leads to a specific structure of the feasible region and, consequently, a distinct optimal solution.

To  investigate these configurations, we introduce the following points that define the extreme limits of the constraint when one of the coefficients is fixed:
\begin{align*}
&\rho_{1,\inf} = \left( \frac{G_2 - G_1}{G_{11} - G_{12}} \right)^2,  \rho_{1,\sup} = \left( \frac{G_2 - G_1 - (G_{21} - G_{22})}{G_{11} - G_{12}} \right)^2,
\\
&\rho_{2,\inf} = \left( \frac{G_2 - G_1}{G_{21} - G_{22}} \right)^2,  \rho_{2,\sup} = \left( \frac{G_2 - G_1 - (G_{11} - G_{12})}{G_{21} - G_{22}} \right)^2.
\end{align*}
These points characterize the extremum values of $ \rho_1 $ when $ \rho_2 = 0 $ and when $ \rho_2 = 1 $, and conversely for $ \rho_2 $. The analysis of these points helps in determining the optimal values of $ \rho_1 $ and $ \rho_2 $ by considering the extreme cases and ensuring feasibility with respect to all constraints.

\begin{figure}[t]
    \centering

    \begin{minipage}[t]{0.24\textwidth}
        \centering
        \includegraphics[width=\textwidth]{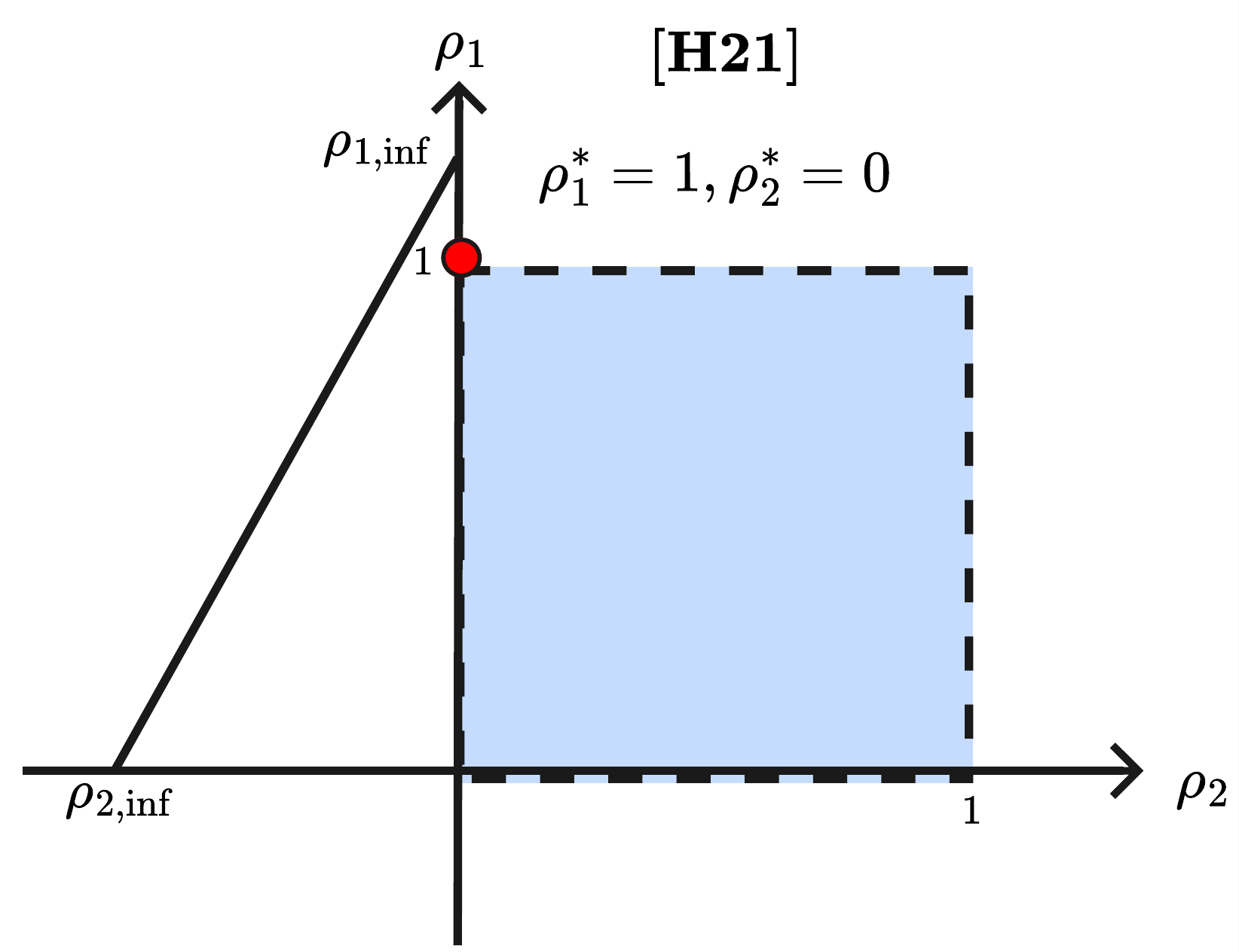}
    \end{minipage}
   
    \vspace{0.4cm}

    \begin{minipage}[t]{0.24\textwidth}
        \centering
        \includegraphics[width=\textwidth]{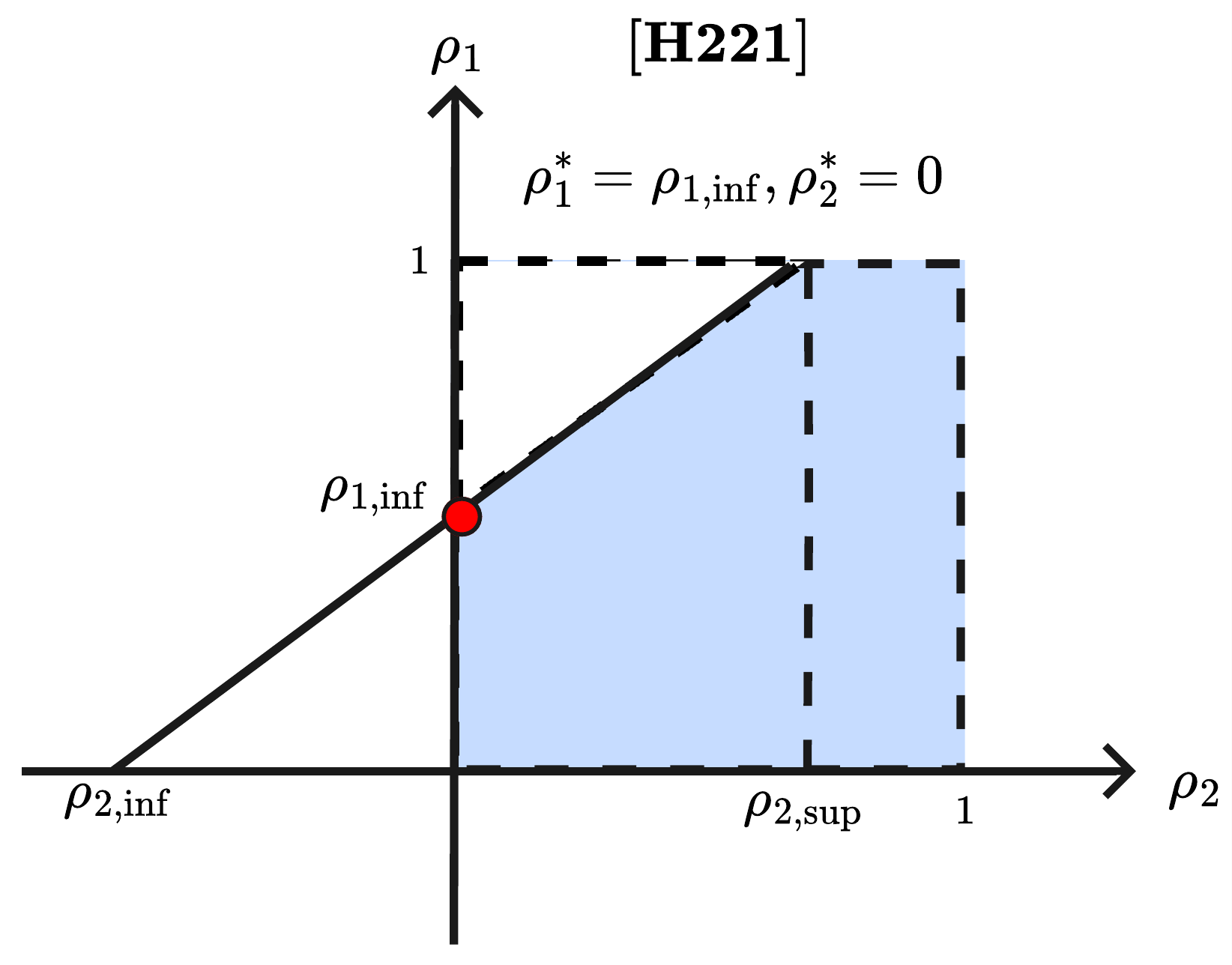}
    \end{minipage}
    \hfill
    \begin{minipage}[t]{0.24\textwidth}
        \centering
        \includegraphics[width=\textwidth]{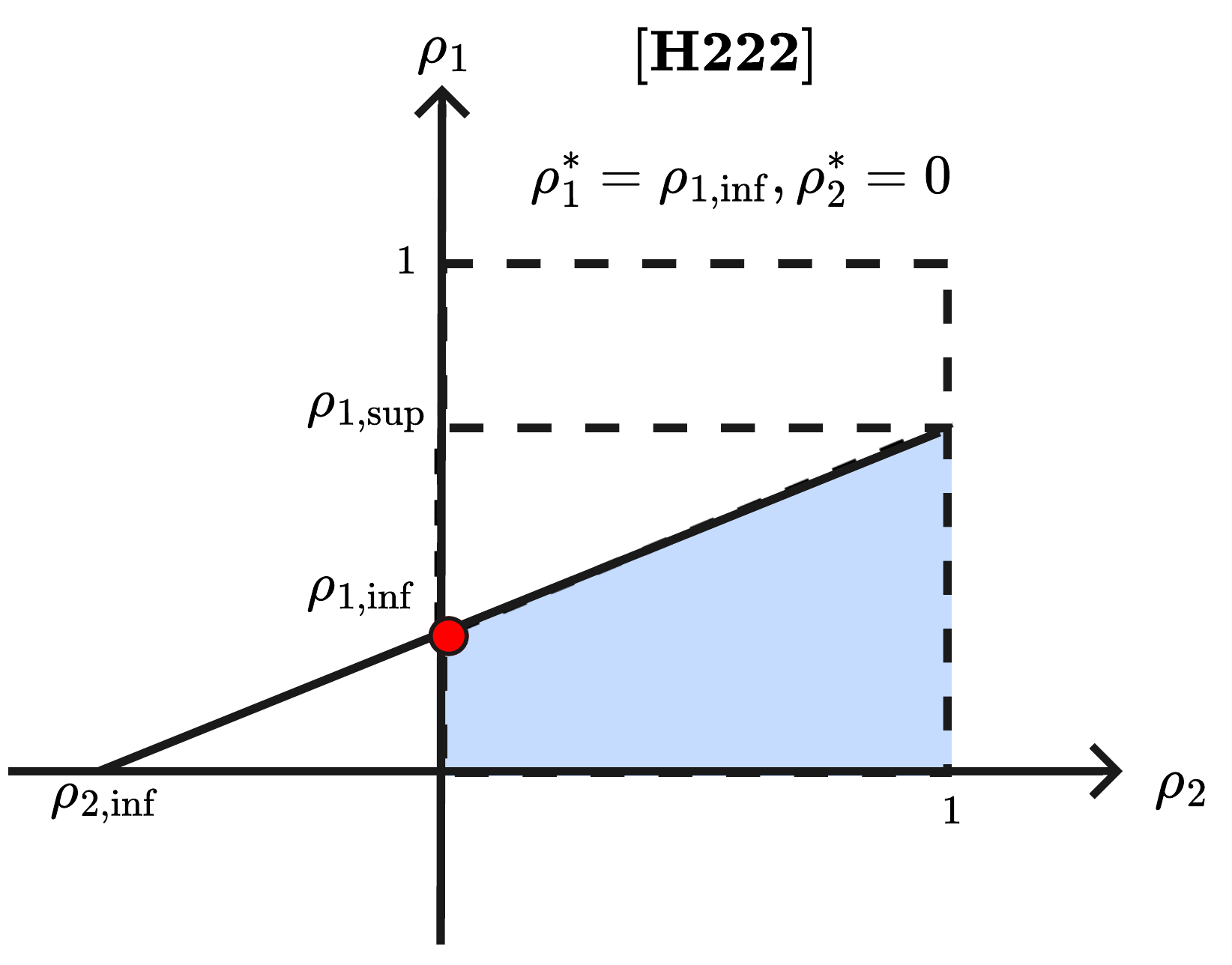}
    \end{minipage}

  \caption{Geometric configuration under case [H2]. The blue area represents the feasible region defined by the constraints, and the red dot marks the corresponding unique Pareto-optimal point.}

    \label{fig:H2_cases}
\end{figure}

\begin{figure}[t]
    \centering

    \begin{minipage}[t]{0.24\textwidth}
        \centering
        \includegraphics[width=\textwidth]{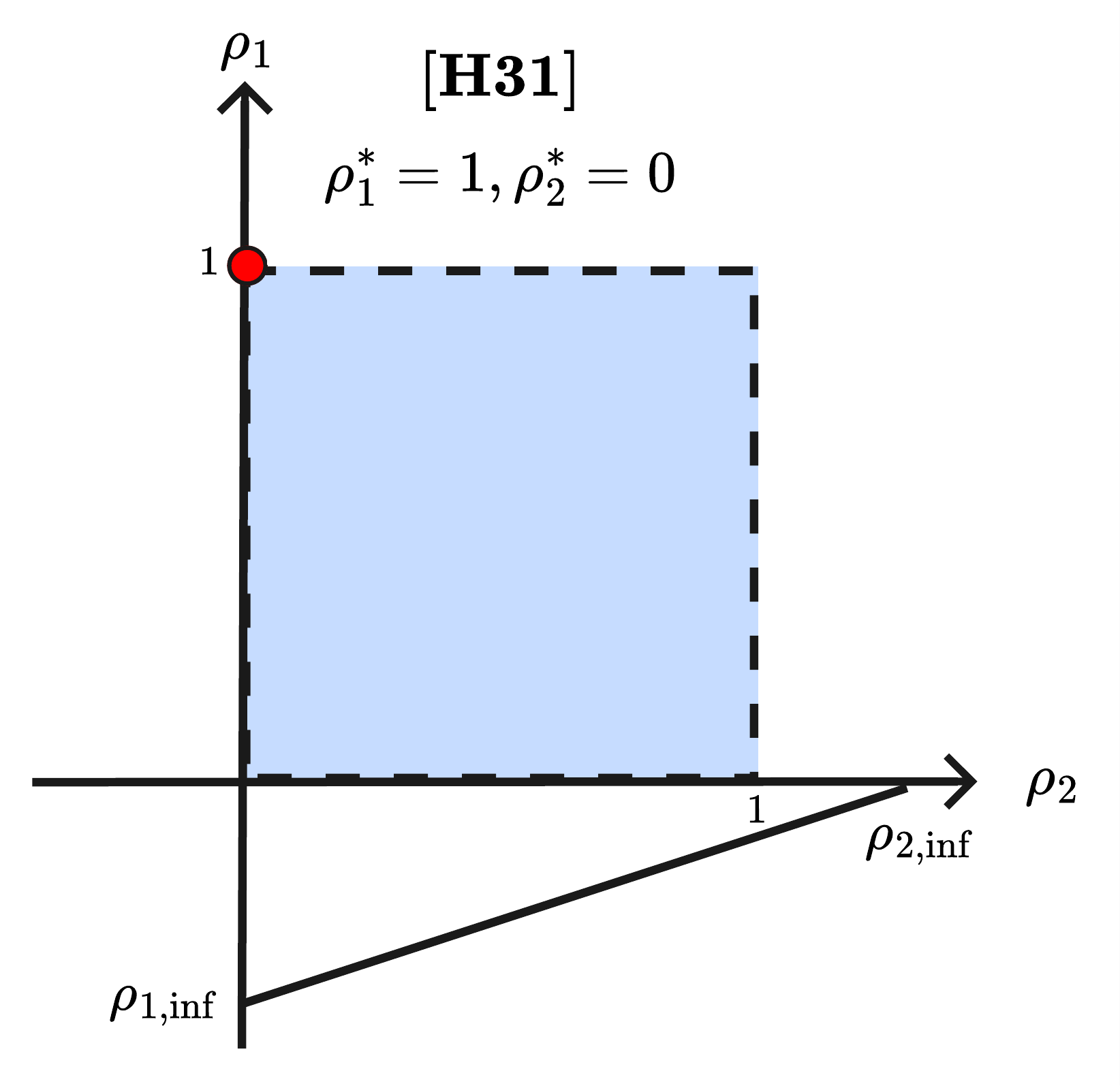}
    \end{minipage}
   
    \vspace{0.4cm}

    \begin{minipage}[t]{0.24\textwidth}
        \centering
        \includegraphics[width=\textwidth]{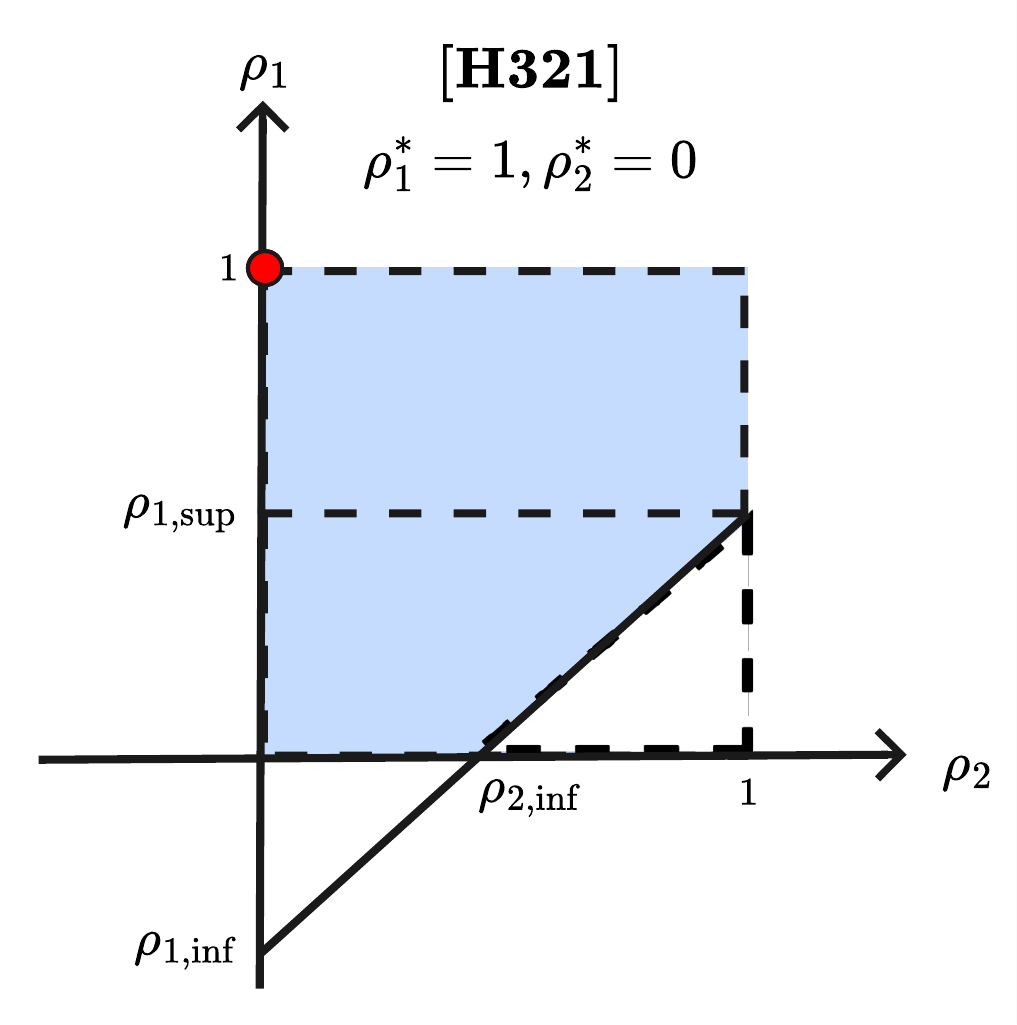}
    \end{minipage}
    \hfill
    \begin{minipage}[t]{0.24\textwidth}
        \centering
        \includegraphics[width=\textwidth]{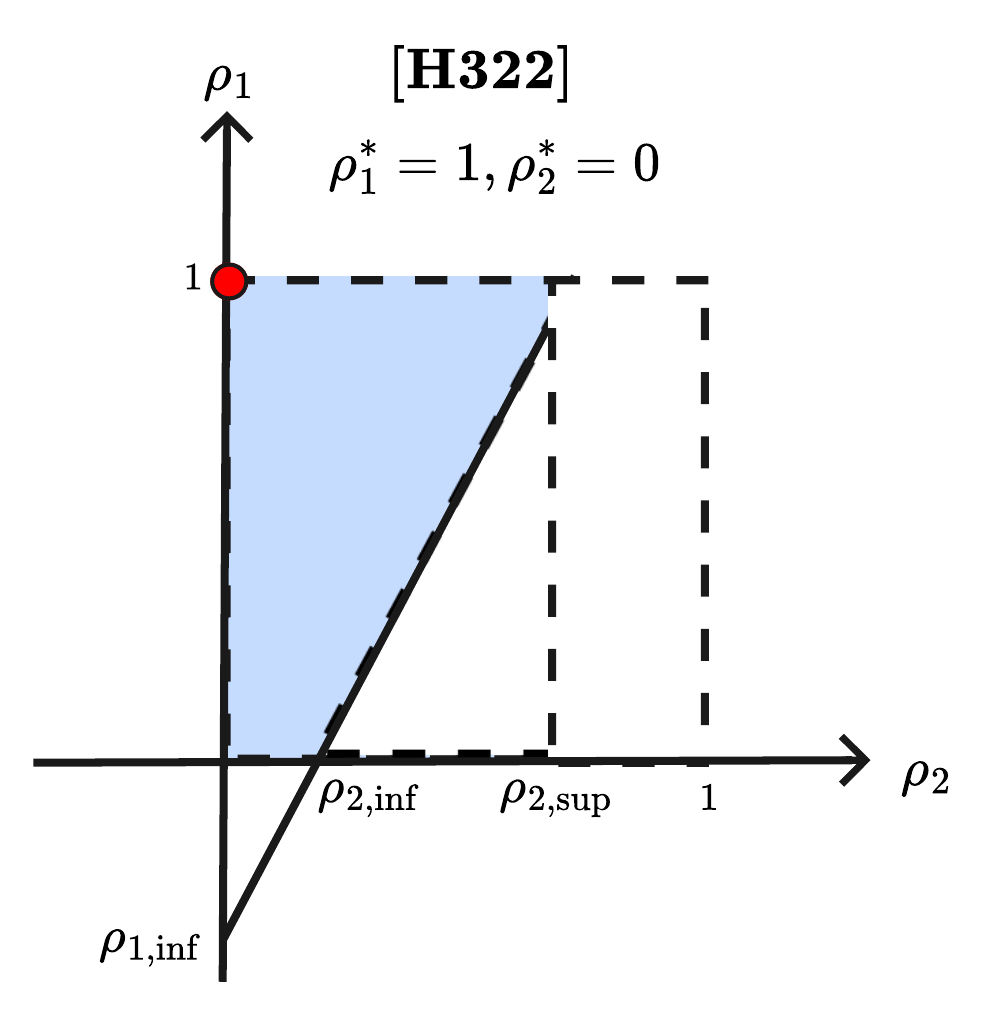}
    \end{minipage}

    \caption{Geometric configuration under case [H3], exhibiting a different structure of the feasible region compared to [H2]. The blue area illustrates the admissible solutions, and the red point identifies the Pareto-optimal solution in this setting.}

    \label{fig:H3_cases}
\end{figure}

We now consider the following scenarios:\\
\textbf{[H1]} If $(G_{11} - G_{12}) \leq 0$ and $(G_{21} - G_{22}) \leq 0$,  the left-hand side of \eqref{Paretoinequality} decreases with both $\rho_1$ and $\rho_2$. Given that the right-hand side $G_2 - G_1$ is assumed to be non-negative, the inequality is always satisfied for all $(\rho_1, \rho_2) \in [0, 1]^2$. Hence, the unique optimal solution is attained at $\rho_1^* = 1$, $\rho_2^* = 0$.\\
\\
\textbf{[H2]} If $(G_{11} - G_{12}) > 0$ and $(G_{21} - G_{22}) \leq 0$, then it follows that  $\rho_{1,\inf}\ge 0$, $\rho_{2,\inf} < 0$ and $\rho_{1,\sup}> \rho_{1,\inf}>0$. The analysis shows that the nature of the optimal solution depends on the value taken by $\rho_{1,\inf}$, giving rise to several distinct sub-cases as shown in Fig.~\ref{fig:H2_cases}.\\
\textbf{[H21]} If $\rho_{1,\inf}>1$ which implies $ \rho_{2,\sup}<0$ and $ \rho_{1,\sup}>1$. In this scenario, the inequality is not satisfied for any $(\rho_1, \rho_2) \in [0,1]^2$, except at the extreme point $(1, 0)$. Hence, the unique Pareto-optimal solution is $\rho_1^* = 1$, $\rho_2^* = 0$.\\
\textbf{[H22]} If $\rho_{1,\inf} \leq 1$ then we have $\rho_{2,\sup}\ge0$ and $\rho_{1,\sup}> \rho_{1,\inf}\ge 0$. This case can be further divided into two sub-cases:
\begin{itemize}
   \item \textbf{[H221]} if $\rho_{1,\sup}>1$, then the inequality remains satisfied for all $\rho_1 \leq 1$ when $\rho_2 = 0$. Thus, the unique optimal solution is $\rho_1^* = \rho_{1,\inf}$, $\rho_2^* = 0$.
   \item \textbf{[H222]} if $\rho_{1,\sup}\leq 1$  then the inequality is satisfied only up to $\rho_1 = \rho_{1,\inf}$ for $\rho_2 = 0$, and the unique Pareto-optimal point becomes $\rho_1^* = \rho_{1,\inf}$, $\rho_2^* = 0$.
\end{itemize}
In summary, the optimal solution in this case is $\rho_1^* = \min\left(1, \rho_{1,\inf}\right)$, $\rho_2^* = 0$.\\
\\
\textbf{[H3]} If $(G_{11} - G_{12}) \leq 0$ and $(G_{21} - G_{22}) > 0$, then we have $\rho_{1,\inf}<0$, $\rho_{2,\inf}\ge0$, $\rho_{2,\sup}>\rho_{2,\inf} \ge 0$. This case is symmetric to [H2], with the roles of $\rho_1$ and $\rho_2$ reversed. Therefore, the analysis follows the same reasoning, leading to similar sub-cases:\\
\textbf{[H31]} If $\rho_{2,\inf}>1$, this implies that $ \rho_{1,\sup}<0$ and $ \rho_{2,\sup}>1$, then the unique Pareto-optimal solution is $\rho_1^* = 1$, $\rho_2^* = 0$.\\
\textbf{[H32]} If $\rho_{2,\inf} \leq 1$ then we have $\rho_{1,\sup}\ge0$ and $\rho_{2,\sup}> \rho_{2,\inf}\ge 0$. This case can be further divided into two sub-cases as shown in Fig.~\ref{fig:H3_cases}.
\begin{itemize}
   \item \textbf{[H321]} if $\rho_{2,\sup}>1$, then the unique optimal solution is $\rho_1^* = 1$, $\rho_2^* = 0$.
   \item \textbf{[H322]} if $\rho_{2,\sup}\leq 1$  then  the unique Pareto-optimal point is still $\rho_1^* = 1$, $\rho_2^* = 0$.
\end{itemize}
Hence, the optimal point in this case is also: $\rho_1^* = 1$, $\rho_2^* = 0$.\\
\\
\textbf{[H4]} If $(G_{11} - G_{12}) > 0$ and $(G_{21} - G_{22}) > 0$, then the left-hand side of \eqref{eq:ParetoBoundary} increases with both $\rho_1$ and $\rho_2$, making the feasible region bounded. The geometric structure of this region has been thoroughly studied in~\cite{el2022energy}, where the feasible boundary was shown to reduce to a linear segment in the $(\rho_1, \rho_2)$ space, parameterized as $\rho_1 = \beta \rho_{1,\inf}$, $\rho_2 = (1 - \beta)\rho_{2,\inf}$ with $\beta \in [0,1]$. Hence, the bi-variable optimization problem under study \textbf{(SEE2)} reduces to the single variable optimization problem:
\begin{align*}
\max_{\beta} SSR(\beta) - \alpha \left( \sum_{k=1}^{2} p_k + P_c \right) , 
\text{ s.t. } 0\leq\beta\leq 1  
\end{align*}
with:
\begin{align*}
SSR(\beta) &\!=\! \frac{1}{2} \log_{2}\!\left(1 + \!p_{2}(G_{2}+ \beta \rho_{1,\inf} G_{12}\!+\! (1-\beta) \rho_{2,\inf}G_{22})^2\!\right)\\
&-\frac{1}{2} \log_{2}\!\left(1 +\! p_{2}(G_{e}+ \beta \rho_{1,\inf} G_{1e}\!+ \!(1-\beta) \rho_{2,\inf}G_{2e})^2\!\right) 
\end{align*}

By analyzing the derivative $\frac{\partial SSR}{\partial \beta}$, and using the initial condition \!\!$\Bigg[\!\!\left( \frac{G_{12}}{G_{1e}} > \frac{G_{22}}{G_{2e}} \!\right) 
\text{ and } 
  \left( \frac{G_{22}}{G_{2e}} \neq \frac{G_2}{G_e}\right) 
     \text{ and }\left( \frac{G_{12}}{G_{1e}} \neq \frac{G_2}{G_e} 
     \right)\!\!\Bigg]$, we find that the objective function is increasing with respect to $\beta$ over its feasible interval $[\beta_{\min}, \beta_{\max}] \subseteq [0,1]$. The exact expressions of these bounds depend on the sub-case considered and are characterized geometrically in~\cite{el2022energy}. As a result, the optimal solution is attained at the upper bound of the feasible segment, i.e., $\beta = \beta_{\max}$. This corresponds in our case to $\rho_1^* = \min(1, \rho_{1,\inf})$ and $\rho_2^* = 0$.\\
\textbf{(\romannumeral 2)} \small$
\left( \frac{G_{12}}{G_{1e}} < \frac{G_{22}}{G_{2e}} \!\right) 
\text{ and } 
  \left( \frac{G_{22}}{G_{2e}} \neq \frac{G_2}{G_e}\right) 
     \text{ and }\left( \frac{G_{12}}{G_{1e}} \neq \frac{G_2}{G_e} 
     \right)$, \normalsize we show that the secrecy sum-rate $SSR(\bm \rho, \mathbf{p} )$ \normalsize is monotonically decreasing w.r.t.  $\rho_1$ and increasing w.r.t. $\rho_2$.  The analysis in this configuration follows, by
symmetry, the same reasoning as in \textbf{(\romannumeral 1)}.

 \end{appendices}

\ifCLASSOPTIONcaptionsoff
  \newpage
\fi

\bibliographystyle{IEEEtran}
\bibliography{IEEEabrv,sample}

\vfill

\end{document}